\documentclass[english]{article}
\usepackage[T1]{fontenc}
\usepackage[latin9]{inputenc}
\usepackage{geometry}
\usepackage{float}
\usepackage{amssymb}
\usepackage{graphicx}
\usepackage{setspace}
\makeatletter
\newcommand{\lyxaddress}[1]{
\par {\raggedright #1
\vspace{1.4em}
\noindent\par}
}

\makeatother

\usepackage{babel}
\begin{document}
\begin{singlespace}

\title{Statistical Properties of the one dimensional Anderson model relevant
for the Nonlinear Schrödinger Equation in a random potential}
\end{singlespace}

\begin{singlespace}

\author{Erez Michaely and Shmuel Fishman}
\end{singlespace}

\maketitle
\begin{singlespace}

\lyxaddress{Physics Department, Technion - Israel Institute of Technology, Haifa
32000, Israel}
\end{singlespace}
\begin{abstract}
\begin{singlespace}
The statistical properties of overlap sums of groups of four eigenfunctions
of the Anderson model for localization as well as combinations of
four eigenenergies are computed. Some of the distributions are found
to be scaling functions, as expected from the scaling theory for localization.
These enable to compute the distributions in regimes that are otherwise
beyond the computational resources. These distributions are of great
importance for the exploration of the Nonlinear Schrödinger Equation
(NLSE) in a random potential since in some explorations the terms
we study are considered as noise and the present work describes its
statistical properties.\end{singlespace}

\end{abstract}
\begin{singlespace}

\section{Introduction}
\end{singlespace}

\begin{singlespace}
In the present work the statistics of the overlap sums and combinations
of the eigenenergies of the one dimensional Anderson model \cite{Anderson1958,Lee1985}
are calculated. These quantities naturally arise in the exploration
of the Nonlinear Schrödinger Equation in a random potential \cite{Fishman2012,Fishman2009a,Krivolapov2010,Kopidakis2008,Flach2009,Skokos2009}.
The overlap sums defined by (\ref{Vnmmm}) measure the overlap of
four eigenfunctions of the Anderson model. The combinations of the
eigenenergies dominate the phase of the nonlinear terms. As a result
of the nonlinearity such terms that affect the dynamics, play a crucial
role in the dynamics of the nonlinear model. Sometimes these terms
are considered as effective noise \cite{Michaely2012,Flach2009,Skokos2009}.
Their statistical properties are crucial for the effective noise theories.
The statistics presented here may be also of interest for mathematicians
exploring the Anderson model and related random models. The Nonlinear
Schrödinger Equation (NLSE) in a random potential takes the form \cite{Fishman2012,Fishman2009a,Krivolapov2010,Flach2009,Skokos2009,Kopidakis2008}
\begin{equation}
i\partial_{t}\psi=H_{0}\psi+\beta\left\vert \psi\right\vert ^{2}\psi,\label{nls}
\end{equation}
 where $H_{0}$ is the linear part with a disordered potential, which
on a lattice takes the form of
\begin{equation}
H_{0}\psi(x)=-\left(\psi(x+1)+\psi(x-1)\right)+\varepsilon(x)\psi(x).\label{eq:linear hamiltonian}
\end{equation}
 In this work it is assumed that $\varepsilon\left(x\right)$ are
identical independent random variables (i.i.d) uniformly distributed
in the interval of $\left[\frac{-W}{2},\frac{W}{2}\right].$ This
is the famous Anderson model \cite{Anderson1958,Lee1985}. Here we
study the model in one dimension where all the states are exponentially
localized \cite{Ishii1973,Lifshits1988}

The NLSE was derived for a variety of physical systems under some
approximations. It was derived in classical optics where $\psi$ is
the electric field by expanding the index of refraction in powers
of the electric field keeping only the leading nonlinear term \cite{Agrawal2007,Schwartz2007,Lahini2008}.
For Bose-Einstein Condensates (BEC), the NLSE is a mean field approximation
where the term proportional to the density $\beta|\psi|^{2}$ approximates
the interaction between the atoms. In this field the NLSE is known
as the Gross-Pitaevskii Equation (GPE) \cite{Dalfovo1999,Pitaevskii2003,Leggett2001,Pitaevskii1961,Gross1961}.

A natural question in this research is whether a wave packet that
is initially localized in space will indefinitely spread for dynamics
controlled by (\ref{nls}) \cite{Fishman2012,Fishman2009a,Krivolapov2010,Flach2009,Skokos2009,Kopidakis2008}. 

It is convenient to expand the wavefunction $\psi$ in the linear
problem eigenfunctions $u_{n}$ 
\begin{equation}
\psi(x,t)=\sum_{n}c_{n}(t)u_{n}(x)e^{-iE_{n}t}\label{eq:expansion of psi}
\end{equation}
 where $E_{n}$ is the eigenenergy of the n-th eigenstate. The width
of the energy spectrum is $\Delta=4+W$ with $E_{n}\in\left[-2-\frac{W}{2},2+\frac{W}{2}\right]$
and $u_{n}$ typically falling off exponentially \cite{Ishii1973,Lifshits1988}
\begin{equation}
u_{n}(x)\approx\frac{e^{-|x_{n}-x|/\xi_{n}}}{\sqrt{\xi_{n}}}\varphi(x)\label{eq:scale of u}
\end{equation}
 where $\varphi(x)$ is a real random function of order unity, $\xi_{n}$
is the localization length for the n-th state and $x_{n}$ is its
localization center. The expansion coefficients $c_{n}(t)$ satisfy
\cite{Fishman2012,Fishman2009a}

\begin{equation}
i\partial_{t}c_{m_{1}}(t)=\beta\sum_{m_{2},m_{3},m_{3}}V_{m_{1}}^{m_{2},m_{3},m_{4}}\exp\left(i\cdot t\Phi_{m_{1}}^{m_{2},m_{3},m_{4}}\right)c_{m_{2}}^{*}c_{m_{3}}c_{m_{4}}\label{Cs}
\end{equation}
 where 
\begin{equation}
V_{m_{1}}^{m_{2},m_{3},m_{4}}=\sum_{x}u_{m_{1}}(x)u_{m_{2}}(x)u_{m_{3}}(x)u_{m_{4}}(x).\label{Vnmmm}
\end{equation}
is the overlap sum and 
\begin{equation}
\Phi_{m_{1}}^{m_{2},m_{3},m_{4}}=E_{m_{1}}+E_{m_{2}}-E_{m_{3}}-E_{m_{4}}\label{eq:phi}
\end{equation}
is the total phase.

The main purpose of this paper it to explore numerically the statistical
properties of the overlap sum $V_{m_{1}}^{m_{2},m_{3},m_{4}}$ and
the total phase $\Phi_{m_{1}}^{m_{2},m_{3},m_{4}}$ as well as of
related quantities. We believe that understanding the qualitative
properties of $V_{m_{1}}^{m_{2},m_{3},m_{4}}$ and $\Phi_{m_{1}}^{m_{2},m_{3},m_{4}}$
will help to build a toy model that may shed light on the spreading
mechanism. Of particular interest will be to find universal distributions
in terms of scaling variables as explained in Sec. 2. These should
be relevant for the entire energy regime. The distributions are found
numerically and are a generalization of the distribution of $V_{0}^{0,0,0}$
that was found analytically by Fyodorov and Mirlin in a narrow energy
range \cite{Mirlin1993}. In Sec. 2 we explore statistical properties
of $V_{m_{1}}^{m_{2},m_{3},m_{4}}$ for weak disorder and in Sec.
3 we will explore statistical properties of $\Phi_{m_{1}}^{m_{2},m_{3},m_{4}}$
for weak disorder. For weak disorder it is expected from the scaling
theory that all statistical properties are determined by the localization
length, while for the strong disorder this does not hold \cite{Shapiro.1988}.
Some results for strong disorder are presented in Sec. 4. The results
are summarized and discussed in Sec. 5.
\end{singlespace}

\begin{singlespace}

\section{Statistical properties of $V_{m_{1}}^{m_{2},m_{3},m_{4}}$ for weak
disorder}
\end{singlespace}

\begin{singlespace}
We explore the values of $V_{m_{1}}^{m_{2},m_{3},m_{4}}$ in the regime
of weak disorder (which corresponds to relatively long localization
length $\xi$). Most of the explorations are numerical. The lattice
size is fixed at $N=500$. For each realization of the $\varepsilon\left(x\right)$,
we computed the eigenfunctions $u_{n}$ and ordered them in space
by the center of norm coordinate, defined by $x_{n}=\sum_{x}x\cdot u_{n}^{2}\left(x\right)$
. We choose $u_{m_{1}}$ with $m_{1}=0$ to be the eigenfunction centered
in the lattice. We studied only the following quantities $V_{0}^{0,0,0}$,
$V_{0}^{0,1,1}$ , $V_{0}^{0,0,1}$,$V_{0}^{0,1,2}$ and $V_{0}^{1,2,3}$
representative of values where $V_{m_{1}}^{m_{2},m_{3},m_{4}}$ are
large. Combinations with $m_{i}\gtrsim\xi$ (taking $m_{1}=0)$ have
negligible values because the overlap sum is a sum of exponentially
decaying functions in space of the form of (\ref{eq:scale of u}).
We calculated these values for $N_{R}=2\cdot10^{4}$ realizations,
and repeated this calculation for 7 disorder strengths in the weak
disorder regime $1\leq W\leq2$ where the maximal localization length
$\xi$ takes the values of $25\lesssim\xi\lesssim103$. In the regime
of weak disorder the maximal localization length is \cite{DerridaB.1984}
\begin{equation}
\xi\approx\frac{96}{W^{2}}.\label{eq:approx xi}
\end{equation}
We computed the distributions of the $V_{m_{1}}^{m_{2},m_{3},m_{4}}$
as follows. We calculated $2\cdot10^{4}$ values of $V_{m_{1}}^{m_{2},m_{3},m_{4}}$
, one for each realization. We know that $\left|V_{m_{1}}^{m_{2},m_{3},m_{4}}\right|$
must satisfy $0<\left|V_{m_{1}}^{m_{2},m_{3},m_{4}}\right|<1$ because
the eigenfunctions are all normalized $\sum_{x}u_{n}^{2}\left(x\right)=1$.
We made a histogram of the values $\left|V_{m_{1}}^{m_{2},m_{3},m_{4}}\right|$
in number of bins $N_{bins}=500$ in the interval $\left[0,1\right]$,
the resulting bin size is $\delta x=0.002$. In order to get the distribution
we normalized the values of the histogram, dividing them by the number
of realizations, $N_{R}$.

In the calculation of the statistical properties of the $V_{m_{1}}^{m_{2},m_{3},m_{4}}$
we distinguish different groups according to the number of different
indices $m_{i}$.
\end{singlespace}

\begin{singlespace}

\subsection{The case $m_{1}=m_{2}=m_{3}=m_{4}=0$ }
\end{singlespace}

\begin{singlespace}
In the case where all indices are equal we have chosen them to be
zero. In this case
\begin{equation}
V_{0}^{0,0,0}=\sum_{x}u_{0}^{4}\left(x\right)\equiv V_{0}.\label{eq:V_0}
\end{equation}
It is just the inverse participation ratio. Its distribution was calculated
analytically by Fyodorov and Mirlin \cite{Mirlin1993} and was found
to satisfy scaling, that is if $P\left(V_{0},\xi\right)$ is the probability
density of $V_{0}$ and the localization length is $\xi$ then, if
one defines a scaling variable 
\begin{equation}
y_{0}=V_{0}\xi\label{eq:scaling IPR}
\end{equation}
its probability density is 
\begin{equation}
P\left(y_{0}\right)=\frac{1}{\xi}P\left(V_{0},\xi\right).\label{eq:scaling IPR2}
\end{equation}
In \cite{Mirlin1993} this scaling was found to hold in a \textit{narrow
range of energy}. In the present work we demonstrate numerically that
it is an excellent approximation also when the maximal localization
length (\ref{eq:approx xi}) is used. The scaling function is different
from the one of \cite{Mirlin1993}.

\begin{figure}
\centering{}\includegraphics[scale=0.5]{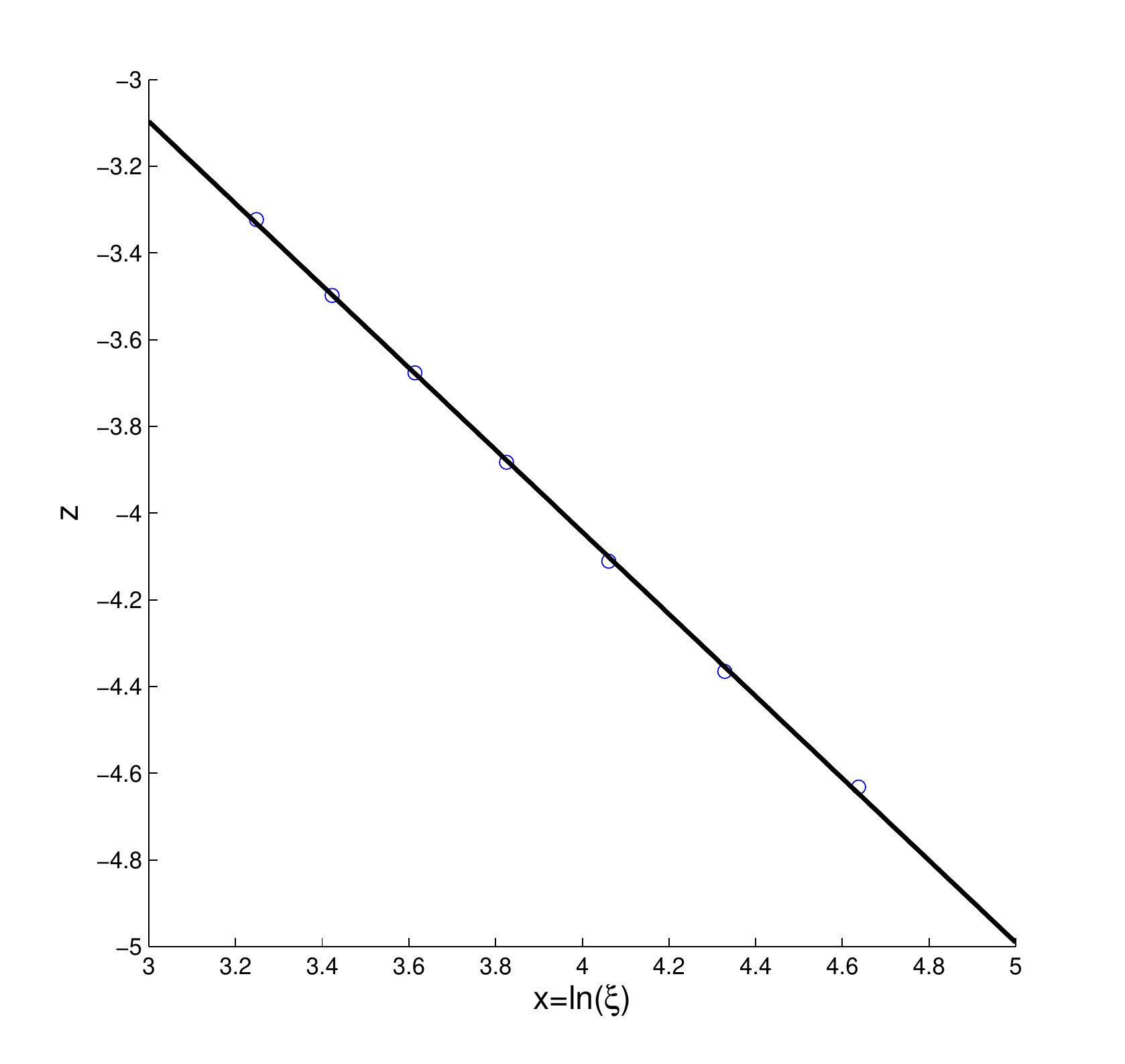}\caption{\label{fig:(a)-(V0000_mean_xi)}a log-log scale of $z=\ln\left(\left\langle V_{0}\right\rangle \right)$
as a function of $x=\ln\left(\xi\right).$ The numerical results are
represented by circles while the line is the best fit. The fit is
$z=-a_{0}x+b_{0}$ for $a_{0}=0.95\ldots$ and $b_{0}=0.26\ldots$. }
\end{figure}

First we verify that the average of $V_{0}$ satisfies 
\begin{equation}
\left\langle V_{0}\right\rangle =\frac{C}{\xi}\label{eq:average V_0-1}
\end{equation}
where C is a constant independant of $\xi$, and $\xi$ is given by
(\ref{eq:approx xi}), as maybe expected from the scaling relation
(in agreement with \cite{Michaely2012}). This is clear from Fig.
\ref{fig:(a)-(V0000_mean_xi)}, and it is found that $C=1.296\ldots$.

\begin{figure}
\includegraphics[scale=0.4]{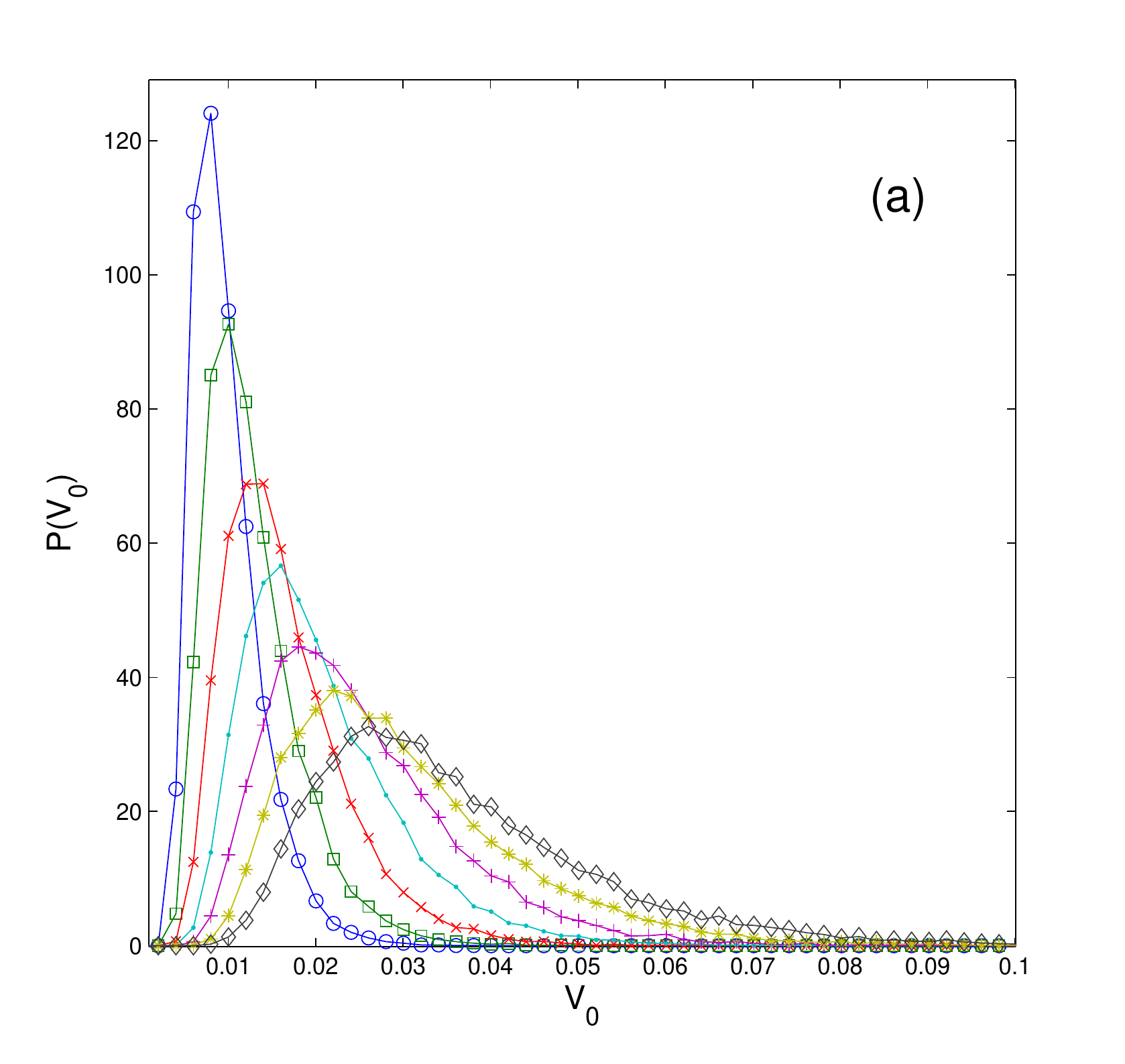}\includegraphics[scale=0.4]{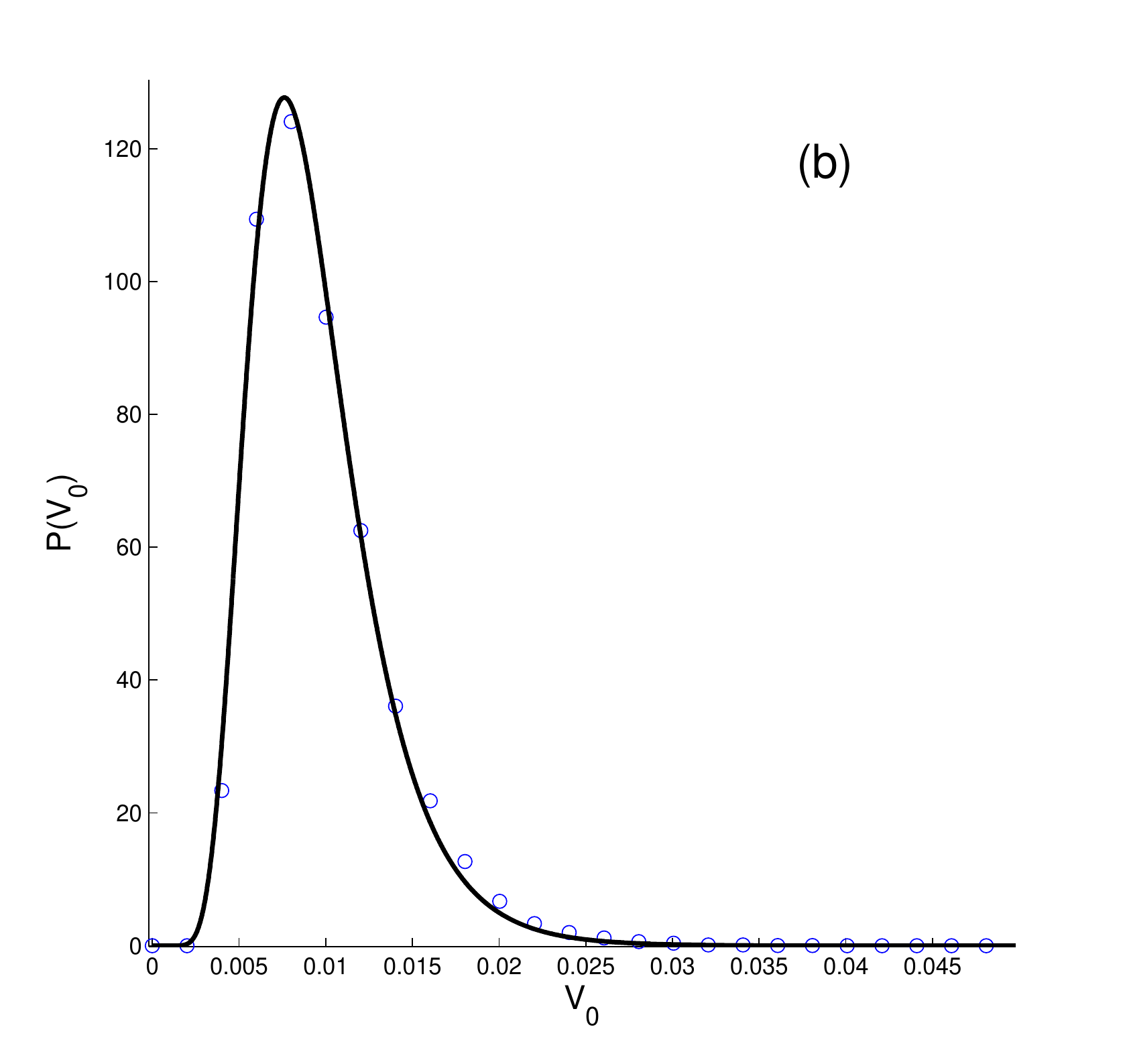}

\includegraphics[scale=0.4]{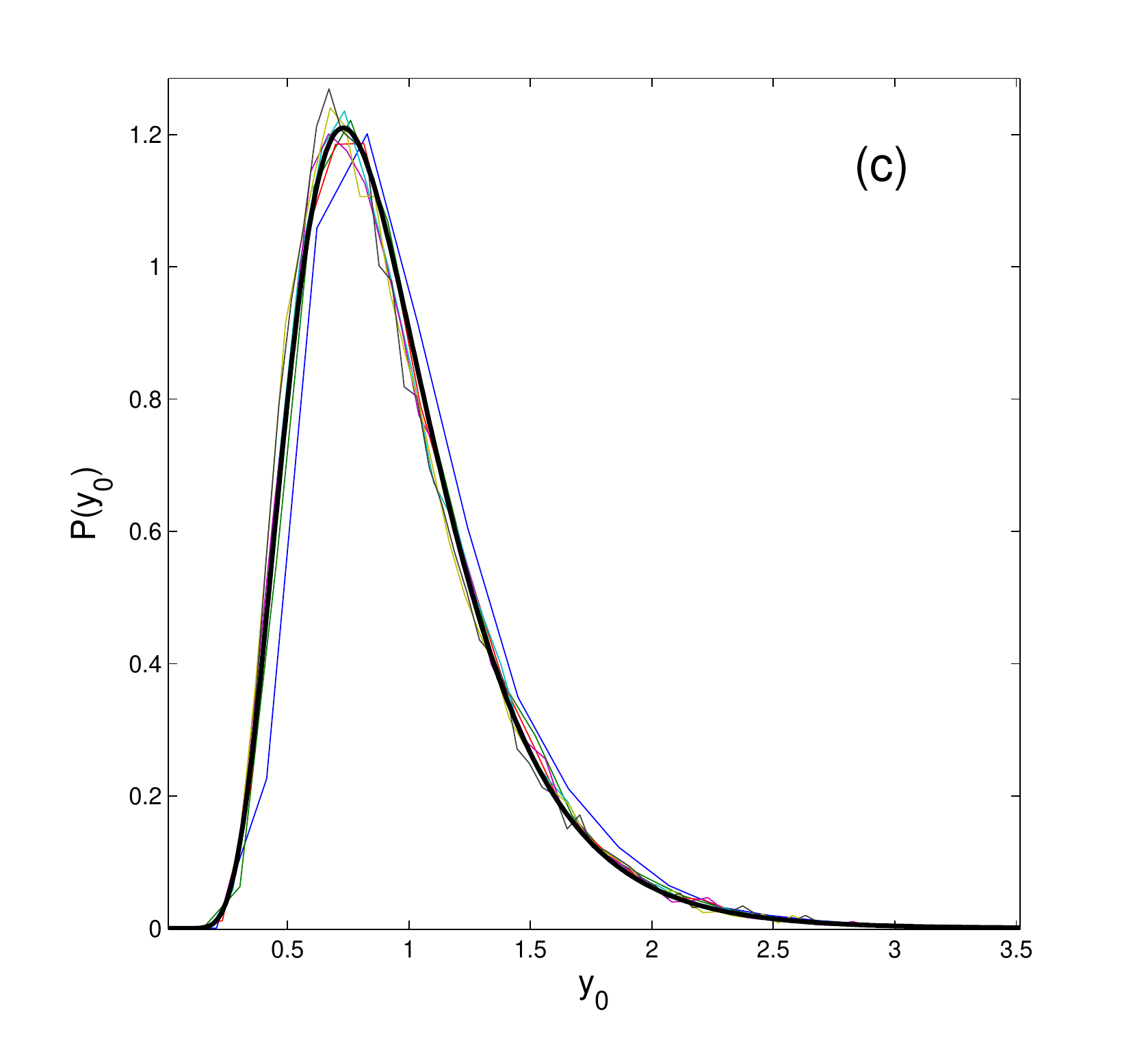}

\caption{\label{fig:distribution IPR}The PDF for various values of weak disorder
strength. (a) The PDF as a function of $V_{0}$ for: blue circles
$W=1$ $\left(\xi\approx103\right)$, green squares $W=1\frac{1}{6}$
$\left(\xi\approx75\right)$ , red crosses $W=1\frac{1}{3}$ $\left(\xi\approx58\right)$,
turquoise dots $W=1\frac{1}{2}$ $\left(\xi\approx46\right)$, purple
pluses $W=1\frac{2}{3}$ $\left(\xi\approx37\right)$, green stars
$W=1\frac{5}{6}$ $\left(\xi\approx30\right)$, black rhombus $W=2$
$\left(\xi\approx25\right)$. (b) The PDF for $W=1$, the results
of the simulation are presented by blue circles, while the solid line
is given by (\ref{eq:scaling of IPR}). (c) $P\left(y_{0}\right)$
as a function of $y_{0}$. The data collapse indicates the scaling
of the PDF with the localization length $\xi$. The colored lines
correspond to various values of the disorder strength. The black thick
line is the best fit for the log normal distribution (\ref{eq:universal_IPR}). }
\end{figure}

The probability density function (PDF) as a function of $V_{0}$ is
presented in Fig. \ref{fig:distribution IPR}a . A typical function
fitted to the numerical data is shown in Fig. \ref{fig:distribution IPR}b
for $W=1$ and it takes the form
\begin{equation}
P\left(V_{0}\right)=c_{1}e^{-\left(c_{2}+c_{3}\cdot\ln\left(V_{0}\right)\right)^{2}}\label{eq:scaling of IPR}
\end{equation}
 with $c_{1}=127.7\ldots$, $c_{2}=9.097\ldots$ and $c_{3}=1.865\ldots$.
We found that the scaling (\ref{eq:scaling IPR}) and (\ref{eq:scaling IPR2})
holds for all weak disorder strengths studied as shown in Fig.\ref{fig:distribution IPR}c.
The resulting scaling function is 
\begin{equation}
P\left(y_{0}\right)=a_{1}e^{-\left(a_{2}+a_{3}\ln\left(y_{0}\right)\right)^{2}}.\label{eq:universal_IPR}
\end{equation}
with $a_{1}=1.21\ldots$, $a_{2}=0.539\ldots$ and $a_{3}=1.71\ldots$.

What is the reason for the scaling? From (\ref{Vnmmm}) and (\ref{eq:scale of u})
it is clear that the magnitude of each of the $u_{m_{i}}$ is of order
$\frac{1}{\sqrt{\xi_{0}}}$ while the number of terms in the sum that
contribute substantially is of order $\xi_{0}$. Therefore $V_{0}$,
although random, it is typically proportional to $\frac{1}{\xi_{0}}$.
Note that  all the contribution to the sum (\ref{Vnmmm}) are positive.

If the calculation is confined to a narrow energy, $\xi_{0}$ is practically
constant and $P\left(y_{0}\right)$ is the function found in \cite{Mirlin1993}.
In the case we study the energy of the site $m_{1}=0$ (middle of
the lattice) varies as the realizations change and an effective average
over the realizations is preformed. Since the density of states (see
Fig. \ref{fig:(a)-The-distribution_weak_enegies}) and the localization
length as a function of energy are flat at the center of the band,
where the localization length is maximal and takes the value close
to (\ref{eq:approx xi}), terms with this value of the localization
length dominate the overlap sum (\ref{Vnmmm}). It is worthwhile to
note that the scaling function (\ref{eq:universal_IPR}) we found
is different from the one found in \cite{Mirlin1993}. It is practically
the average of the function found in \cite{Mirlin1993} over energy.

Now we consider the cases where the $m_{i}$ take two different values
say $V_{0}^{0,1,1}$ of $V_{0}^{0,0,1}$.
\end{singlespace}

\begin{singlespace}

\subsection{Distribution of $V_{0}^{0,1,1}$}
\end{singlespace}

\begin{figure}
\includegraphics[scale=0.5]{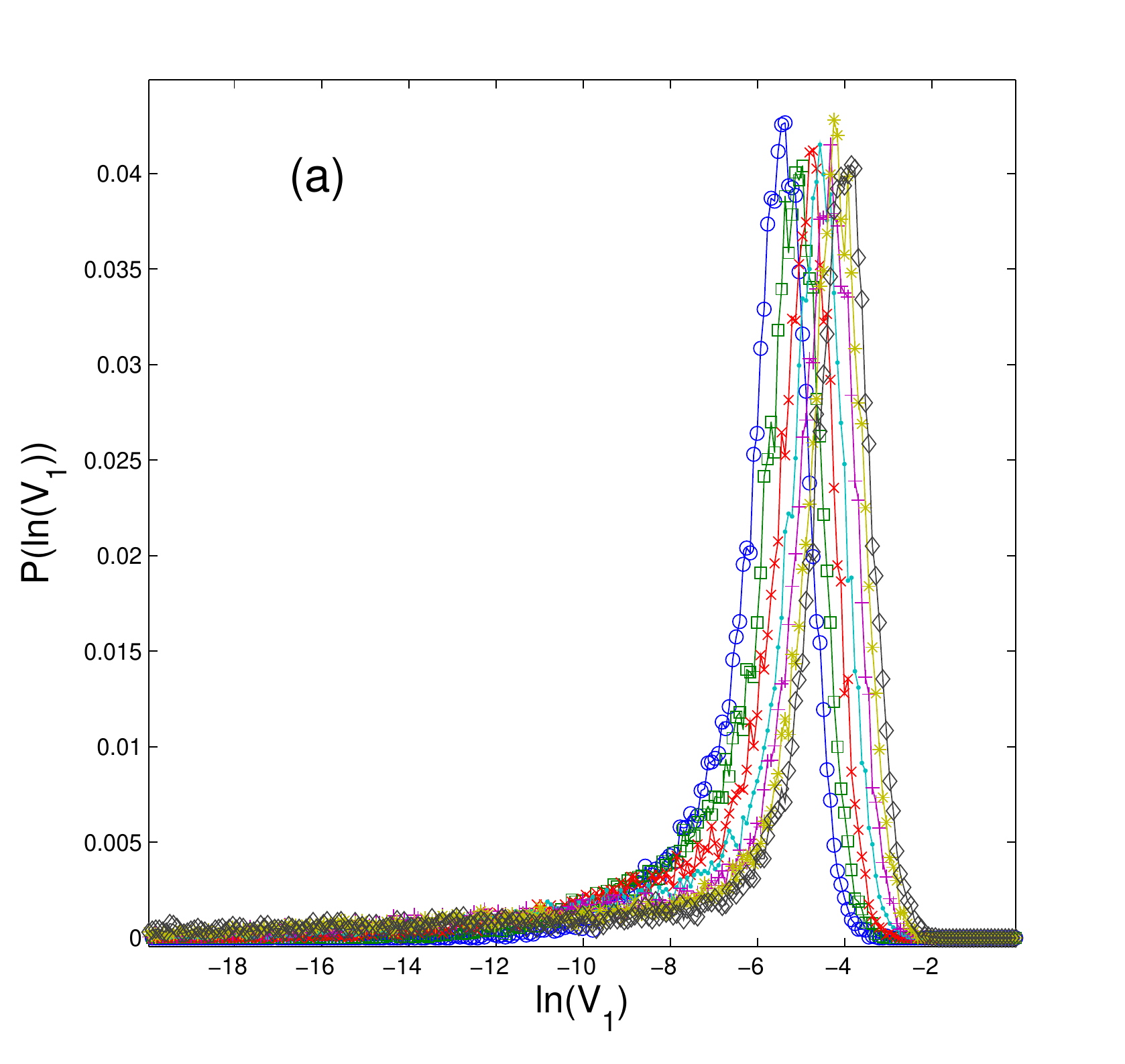}\includegraphics[scale=0.5]{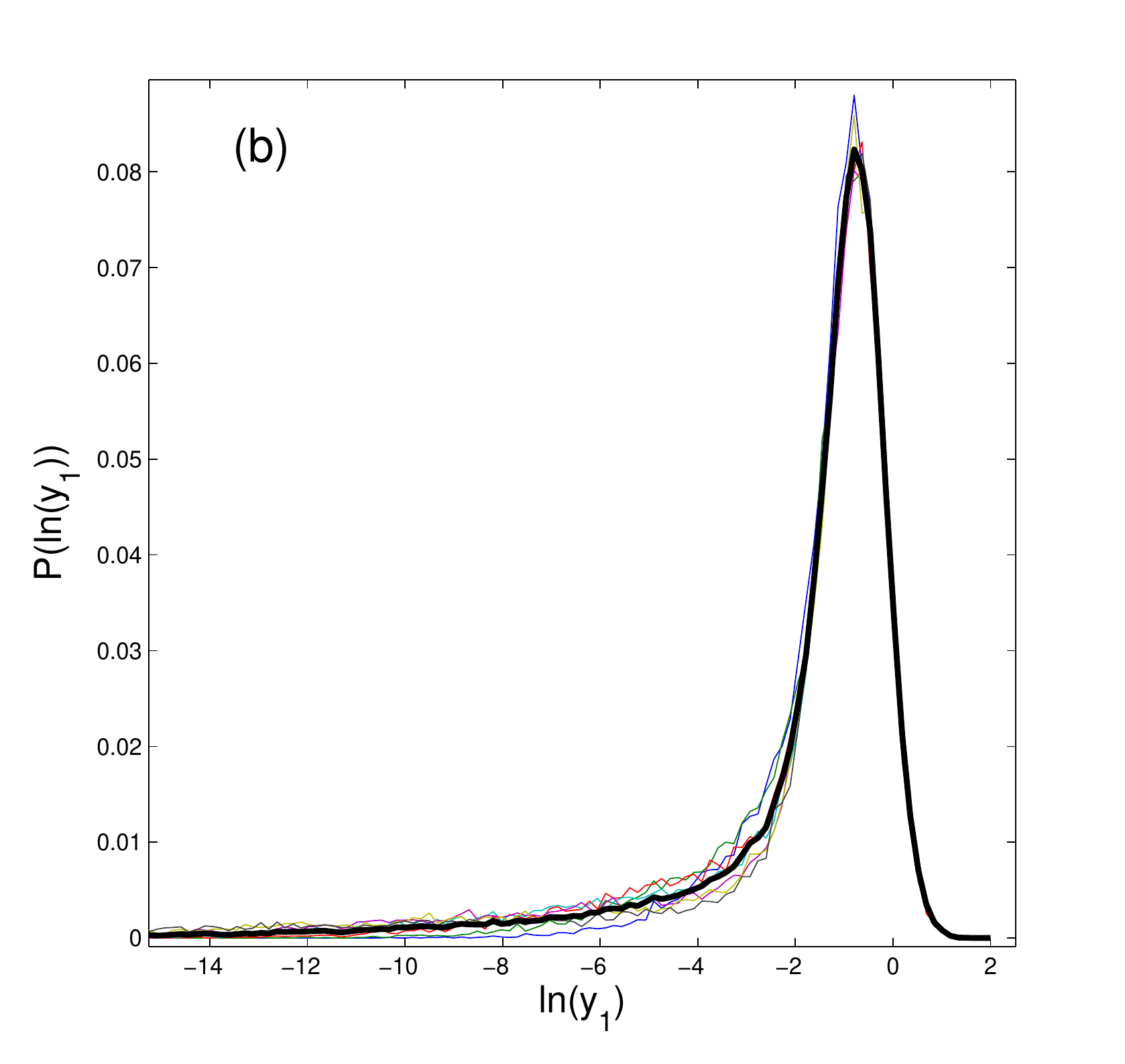}

\caption{\label{fig:distribution_ln_nnn1n1} The PDF of $\ln\left(V_{1}\right)$
for various values of weak disorder strength presented in Fig \ref{fig:distribution IPR}.
(a) The PDF as a function of $\ln\left(V_{1}\right)$ (b) The scaled
PDF as a function of $\ln\left(y_{1}\right)$ the black thick line
represents the best fit. The data collapse indicates the scaling of
the PDF with the localization length $\xi$. The colored lines correspond
to various strengths of disorder. }
\end{figure}

\begin{singlespace}
Also here an argument similar to the one presented in the previous
section holds, but the localization lengths of the two wave functions
involved are different,the overlap sum is of the order $\frac{1}{\xi_{0}+\xi_{1}}$,
therefore $V_{0}^{0,1,1}$ behaves as $\frac{1}{\xi}.$ In order to
investigate the distribution of $V_{0}^{0,1,1},$ denoted here by
$V_{1},$ which consists of many near zero values, we generated the
histogram of $\ln\left(V_{1}\right)$ rather than $V_{1}$ . In Fig.
\ref{fig:distribution_ln_nnn1n1} we present the distribution of $P\left(\ln\left(V_{1}\right)\right)$
as a function of $\ln\left(V_{1}\right)$. The best fit for the scaling
function, in terms of the scaling variable 
\begin{equation}
y_{1}=V_{1}\xi\label{eq:scaling V_1}
\end{equation}
 is shown there as well. As expected $\left\langle V_{1}\right\rangle $
satisfies a relation similar to (\ref{eq:average V_0-1}) but with
$C=0.429\ldots$ (in agreement with \cite{Michaely2012}).
\end{singlespace}

\begin{singlespace}

\subsection{Distribution of $V_{0}^{0,0,1}$}
\end{singlespace}

\begin{figure}
\includegraphics[scale=0.5]{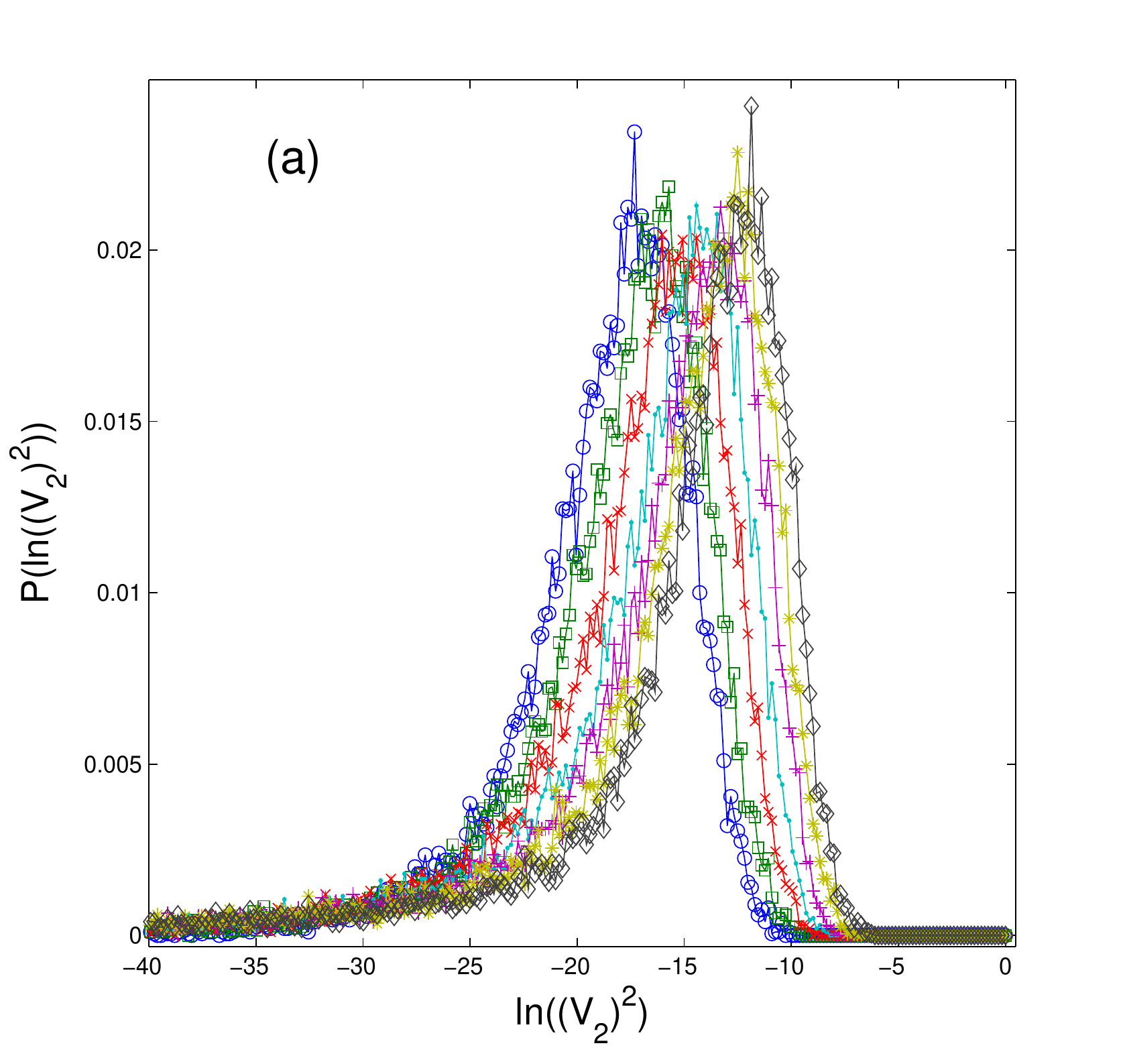}\includegraphics[scale=0.5]{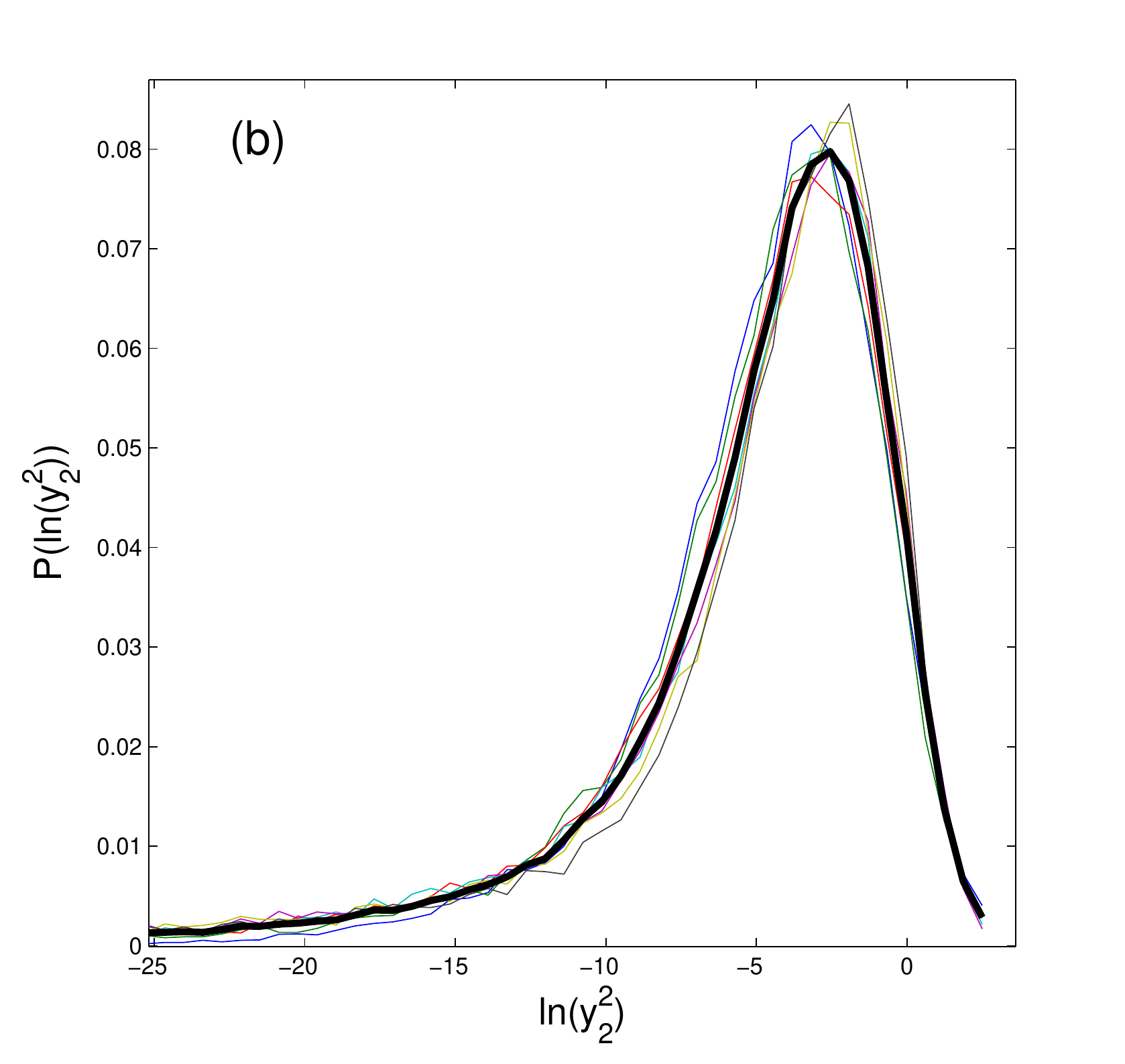}

\caption{\label{fig:distribution_nnn1} As in Fig. \ref{fig:distribution_ln_nnn1n1}
the PDF of $\ln\left(V_{2}^{2}\right)$ for various values of weak
disorder strength presented there. (a) The PDF as a function of $\ln\left(V_{2}^{2}\right).$
(b) The scaled PDF as a function of $\ln\left(y_{2}^{2}\right)$.
The black thick line is the best fit. The number of bins here is $200$.}
\end{figure}

\begin{singlespace}
Let us denote $V_{0}^{0,0,1}\equiv V_{2}$. From the definition (\ref{Vnmmm})
it is clear that $\left\langle V_{2}\right\rangle =0$. Therefore
to estimate the typical value of $V_{2}$ we study $\left\langle V_{2}^{2}\right\rangle $.
It can be estimated by 
\begin{equation}
\left\langle V_{2}^{2}\right\rangle =\left\langle \left(\sum_{x}u_{0}^{3}\left(x\right)u_{1}\left(x\right)\right)^{2}\right\rangle \thickapprox\left\langle \left(\sum_{x}u_{0}^{6}\left(x\right)u_{1}^{2}\left(x\right)\right)\right\rangle .
\end{equation}
 It is of order $\frac{1}{\xi_{0}^{2}\left(3\xi_{1}+\xi_{0}\right)}$.
Therefore it is reasonable that $\left\langle V_{2}^{2}\right\rangle \thicksim\frac{1}{\xi^{3}}$.
Indeed one finds
\begin{equation}
\sqrt{\left\langle V_{2}^{2}\right\rangle }=\bar{C}\xi^{-1.5}
\end{equation}
with $\bar{C}=0.566\ldots$ independent of $\xi$. This motivates
us to introduce the scaling variable 
\begin{equation}
y_{2}=V_{2}\xi^{1.5}.\label{eq:scaling variable y_2}
\end{equation}
In Fig. \ref{fig:distribution_nnn1} we show the distribution of $P\left(\ln\left(V_{2}^{2}\right)\right)$
as a function of $\ln\left(V_{2}^{2}\right)$ in terms of the scaling
variable $y_{2}$. 
\end{singlespace}

\begin{singlespace}

\subsection{Distribution of $V_{m_{1}}^{m_{2},m_{3},m_{4}}$ when 3 or 4 different
$m_{i}$ are involved}
\end{singlespace}

\begin{figure}
\includegraphics[scale=0.5]{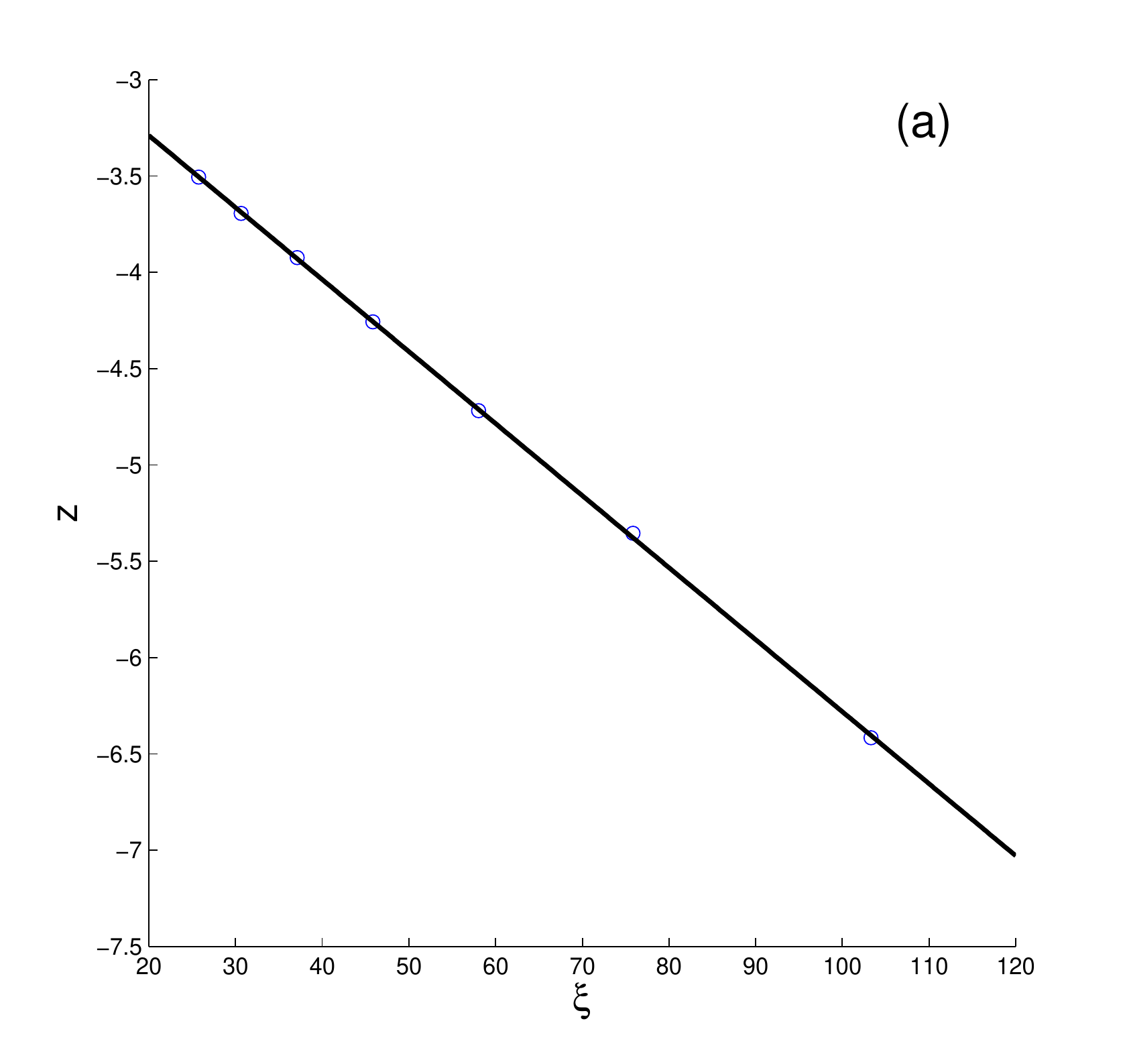}\includegraphics[scale=0.5]{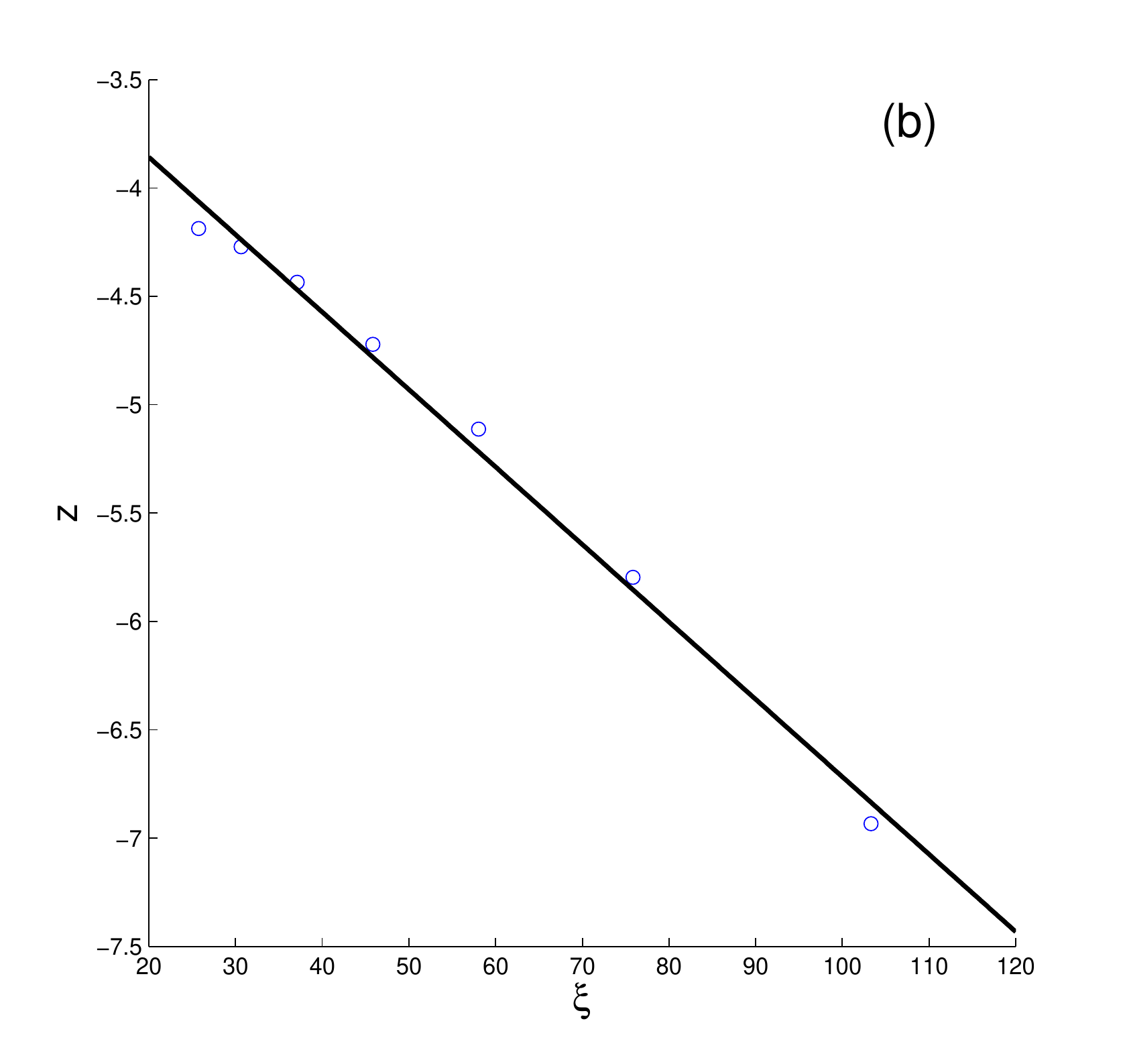}

\caption{\label{fig:(a) mean-xi_3 indices}(a) $z=\ln\sqrt{\left\langle \left(V_{0}^{0,1,2}\right)^{2}\right\rangle }$
as a function of $\xi$. The fit is $z=-0.037\cdot\xi-2.5$ (b) $z=\ln\sqrt{\left\langle \left(V_{0}^{1,2,3}\right)^{2}\right\rangle }$
as a function of $\xi$. The fit is $z=-0.036\cdot\xi-3.1$}
\end{figure}

\begin{singlespace}
In this case we could not find any simple scaling relation. The averages
are found to be exponential in $\xi$, as one can see from Fig. \ref{fig:(a) mean-xi_3 indices}

\end{singlespace}

\begin{singlespace}

\section{Statistical properties of $\Phi_{n}^{m_{1},m_{2},m_{3}}$ for weak
disorder}
\end{singlespace}

\begin{singlespace}
In this section we explore the statistical properties of $\Phi$ defined
in Sec. 1 and the distribution of the eigenenergies $E_{n}.$ For
the weak disorder regime we fixed the lattice size $N=500$ and computed
the eigenenergies for $N_{R}=10^{3}$ realizations. We repeated this
calculation for 7 different values of the disorder strength ,$1<W<2$
which correspond to $25\lesssim\xi\lesssim103$. 
\end{singlespace}

\begin{singlespace}

\subsection{Distribution of $E_{n}$ }
\end{singlespace}

\begin{figure}
\includegraphics[scale=0.5]{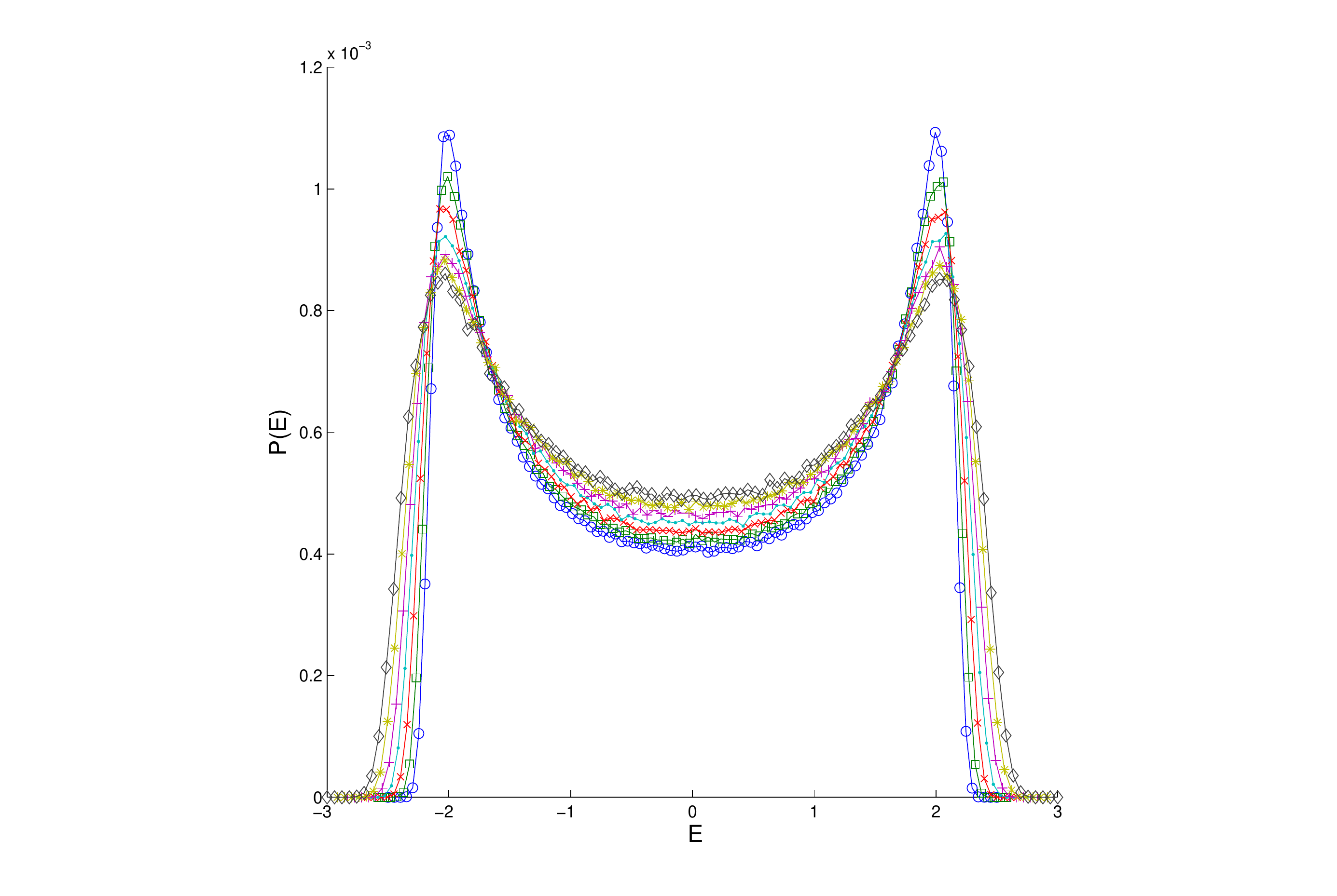}\caption{\label{fig:(a)-The-distribution_weak_enegies} The PDF of $E_{n}$
for various weak disorder strength. blue circles $W=1$ $\left(\xi\approx103\right)$,
green squares $W=1\frac{1}{6}$ $\left(\xi\approx75\right)$ , red
crosses $W=1\frac{1}{3}$ $\left(\xi\approx58\right)$, turquoise
dots $W=1\frac{1}{2}$ $\left(\xi\approx46\right)$, purple pluses
$W=1\frac{2}{3}$ $\left(\xi\approx37\right)$, green stars $W=1\frac{5}{6}$
$\left(\xi\approx30\right)$, black rhombus $W=2$ $\left(\xi\approx25\right)$. }
\end{figure}

\begin{singlespace}
The distribution of the eigenenergies for the weak disorder regime,
as plotted in Fig. \ref{fig:(a)-The-distribution_weak_enegies} is
symmetric around $E=0$ and characterized by convex function in the
middle and sharply decaying function at the boundaries.

\end{singlespace}

\begin{singlespace}

\subsection{Distribution of $\Phi^{+}\equiv E_{n}+E_{m}$ }
\end{singlespace}

\begin{figure}
\includegraphics[scale=0.5]{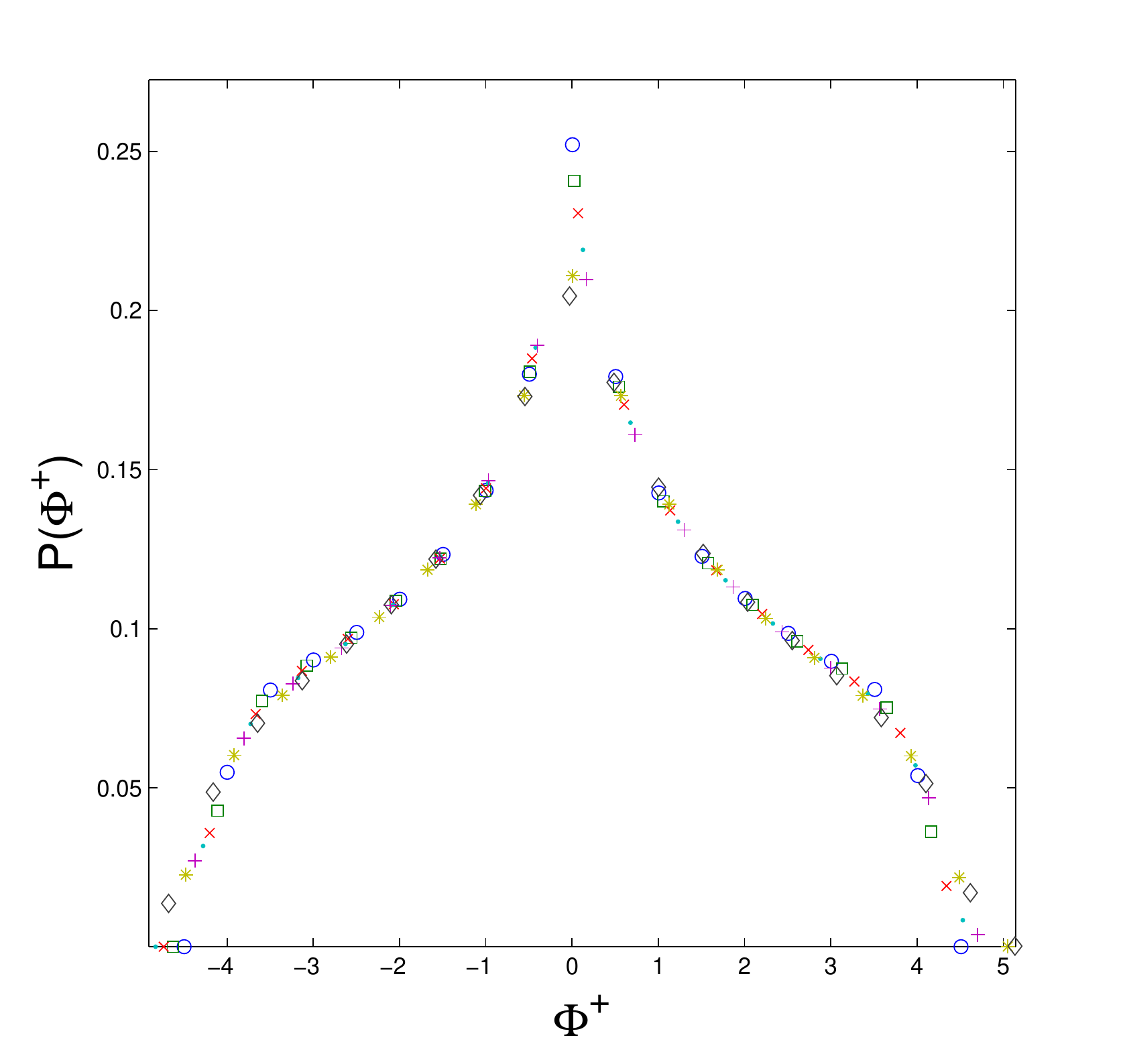}

\caption{\label{fig:sum_distirnution_weak}The distribution of $\Phi^{+}$
for strengths of disorder as in Fig. \ref{fig:(a)-The-distribution_weak_enegies}
using the same symbols.}

\end{figure}

\begin{singlespace}
We calculated the distribution of the sums of two eigenenergies obtained
for the same realization. The motivation for calculation of these
sums is from the terms where $m_{1}=m_{2}=0$ and $m_{3},m_{4}$ are
arbitrary in (\ref{Cs}). In Fig. \ref{fig:sum_distirnution_weak}
we plot distributions of $\Phi^{+}$ with various disorder strengths.
Note the maximal value of the distribution decreases with $W$. We
found the following relation between the maximal value of the distribution
which we will denote $\Phi_{m}$ and $\xi$, 
\begin{equation}
\Phi_{m}\left(\xi\right)=0.13\xi^{0.15}
\end{equation}

\end{singlespace}

\begin{singlespace}

\subsection{Distribution of $\Phi^{-}\equiv E_{n}-E_{m}$ }
\end{singlespace}

\begin{figure}[H]
\includegraphics[scale=0.5]{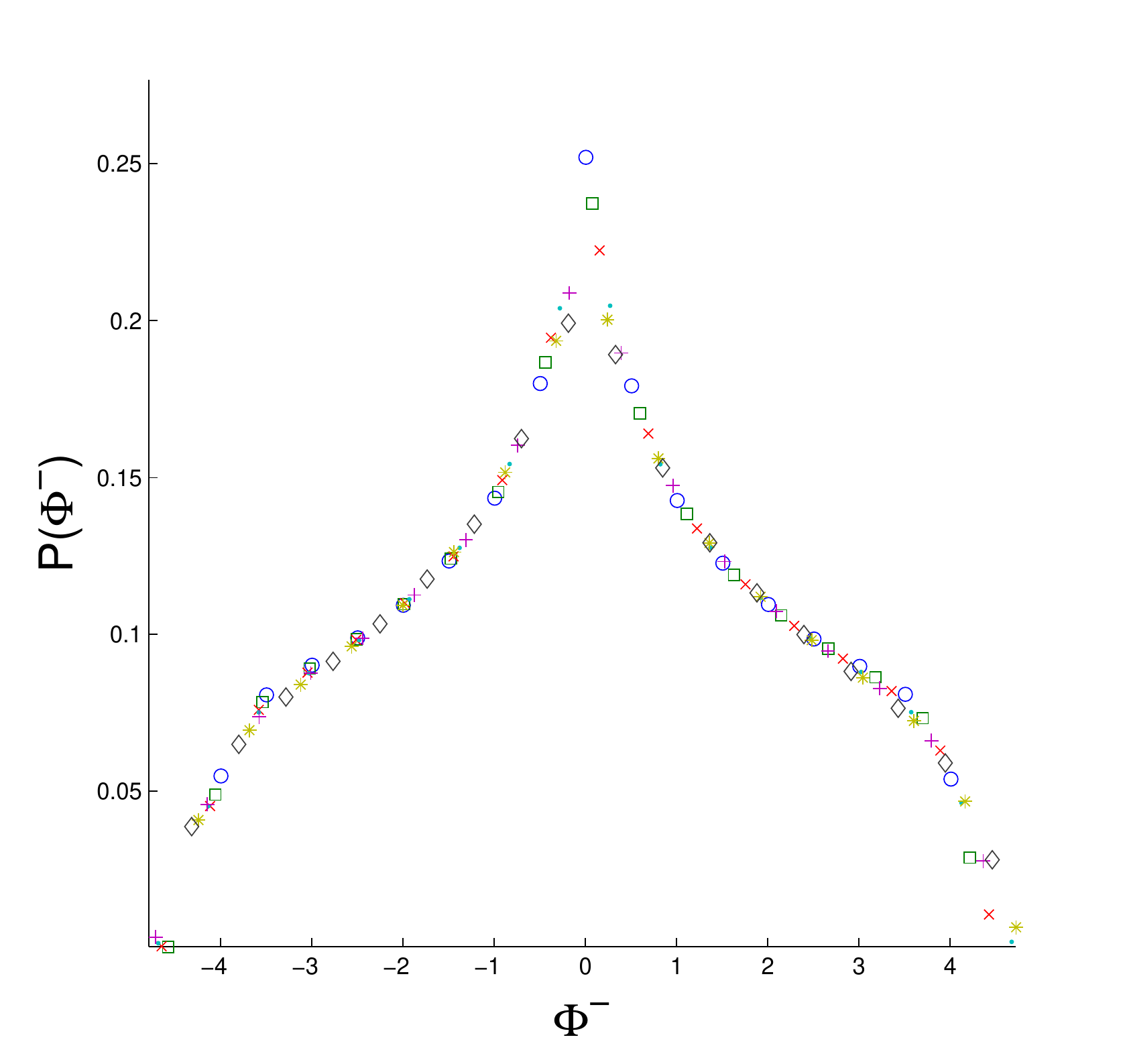}\caption{\label{fig:diff_distribution_weak} (a) The distribution of $\Phi^{-}$
for strengths of disorder as in Fig. \ref{fig:(a)-The-distribution_weak_enegies}.
Number of bins used is $100.$}
\end{figure}

\begin{figure}[H]
\includegraphics[scale=0.5]{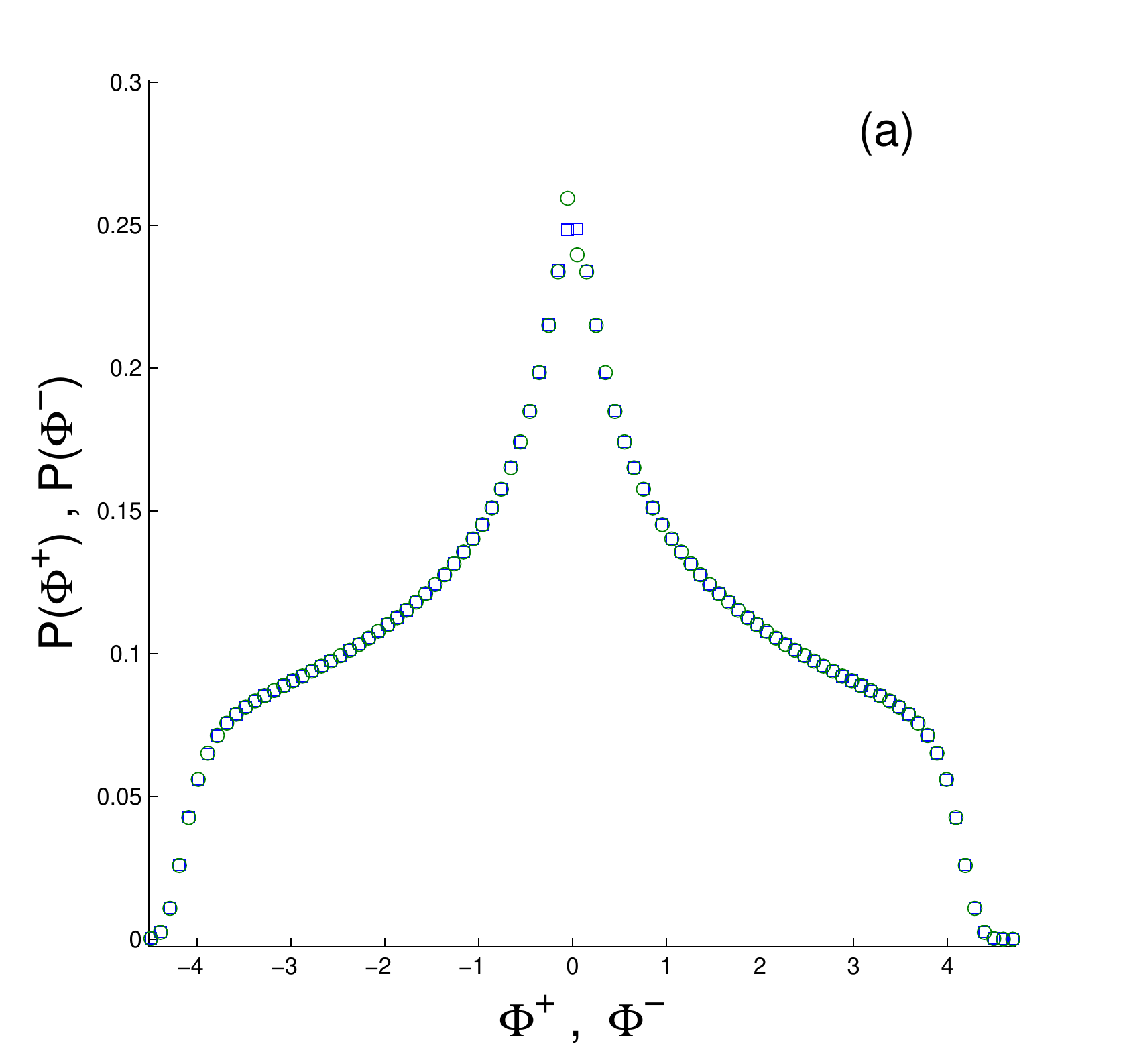}\includegraphics[scale=0.5]{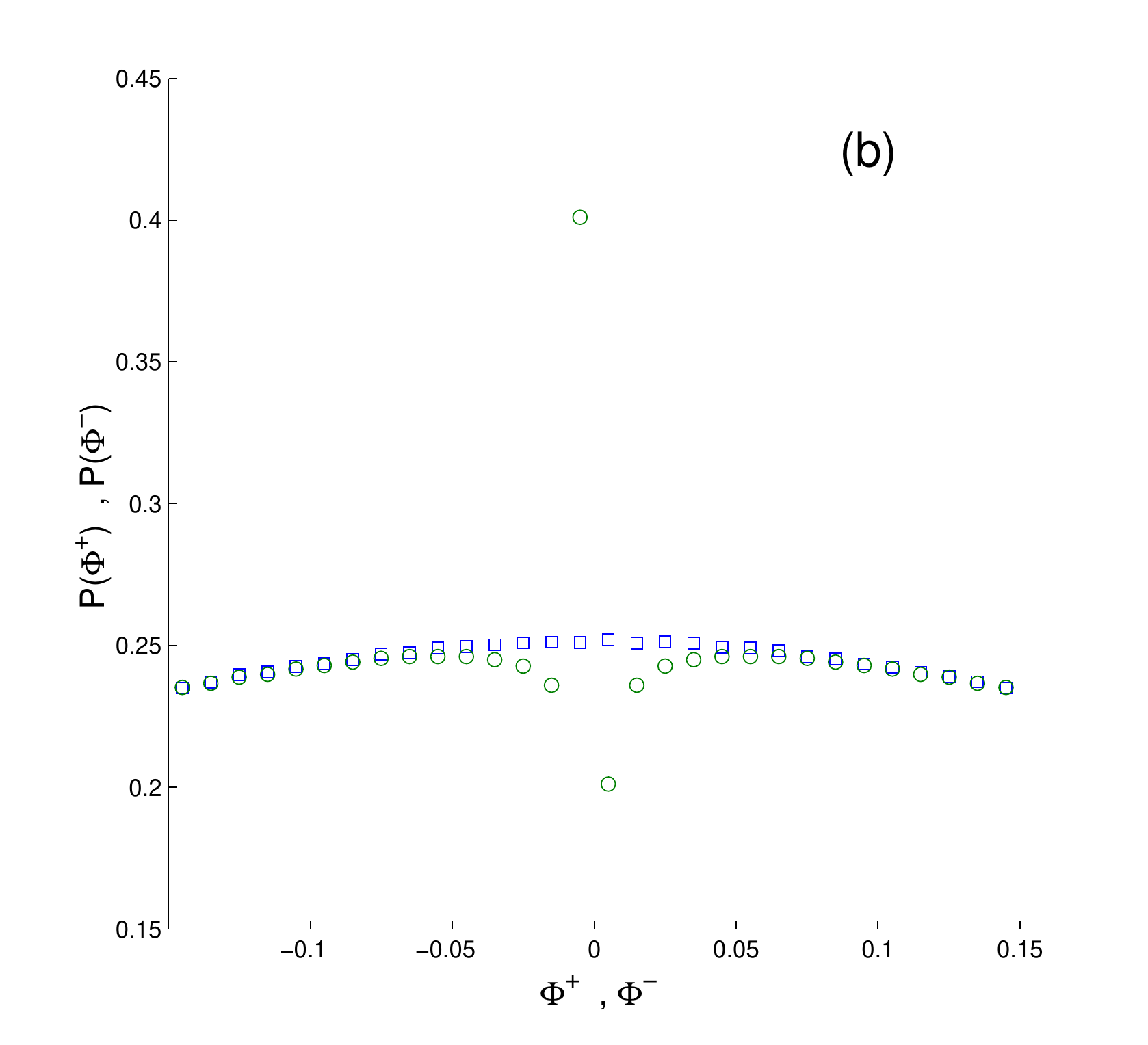}

\caption{\label{fig:sum and diff together}(a) The PDF of $\Phi^{+}$ (blue
squares) and $\Phi^{-}$ (green circles) for $W=1$, total number
of bins used is $10^{2}$. (b) zoom of (a) around $\Phi^{+}=0$ and
$\Phi^{-}=0$, total number of bins used is $10^{3}$.}

\end{figure}

\begin{singlespace}
We calculated the distribution of the differences between two eigenenergies
obtained for the same realization. Despite the symmetric nature of
the distribution of $E_{i}$ the value of $\Phi^{-}$ differs from
$\Phi^{+}$ because of level repulsion \cite{Veksler2010,Rivkind2011}.
Therefore we anticipate a relatively large peak at $\Phi^{-}=0$,
as shown if Fig. \ref{fig:diff_distribution_weak}. A comparison between
$\Phi^{-}$ and $\Phi^{+}$ is presented in Fig. \ref{fig:sum and diff together}.
One can see that, the distributions differ substantially only near
$\Phi^{-}=0$ and $\Phi^{+}=0$ because of level repulsion.

\end{singlespace}

\begin{singlespace}

\subsection{Distribution of $\Phi_{m_{1}}^{m_{2},m_{3},m_{4}}$ }
\end{singlespace}

\begin{figure}[H]
\includegraphics[scale=0.5]{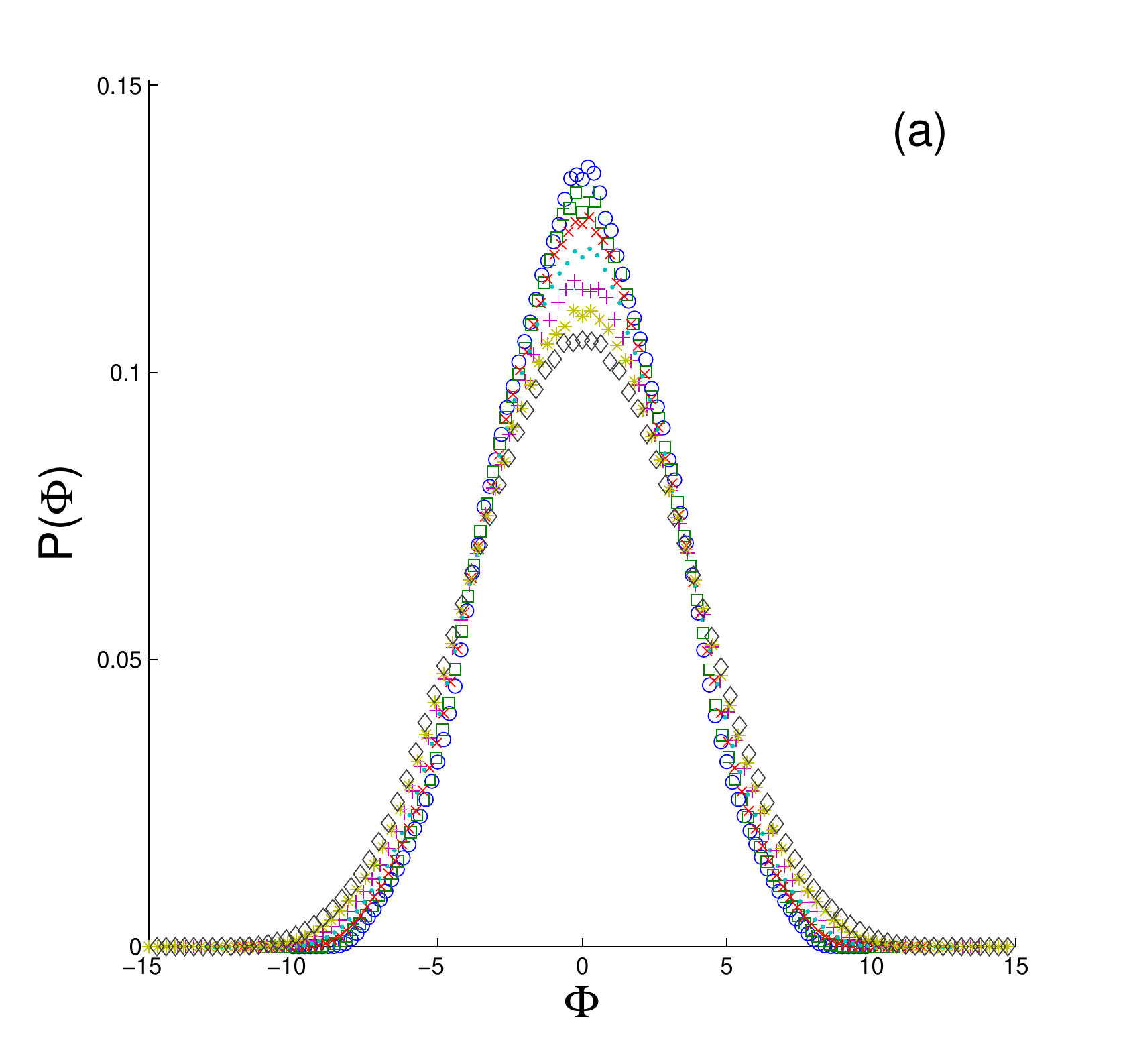}\includegraphics[scale=0.5]{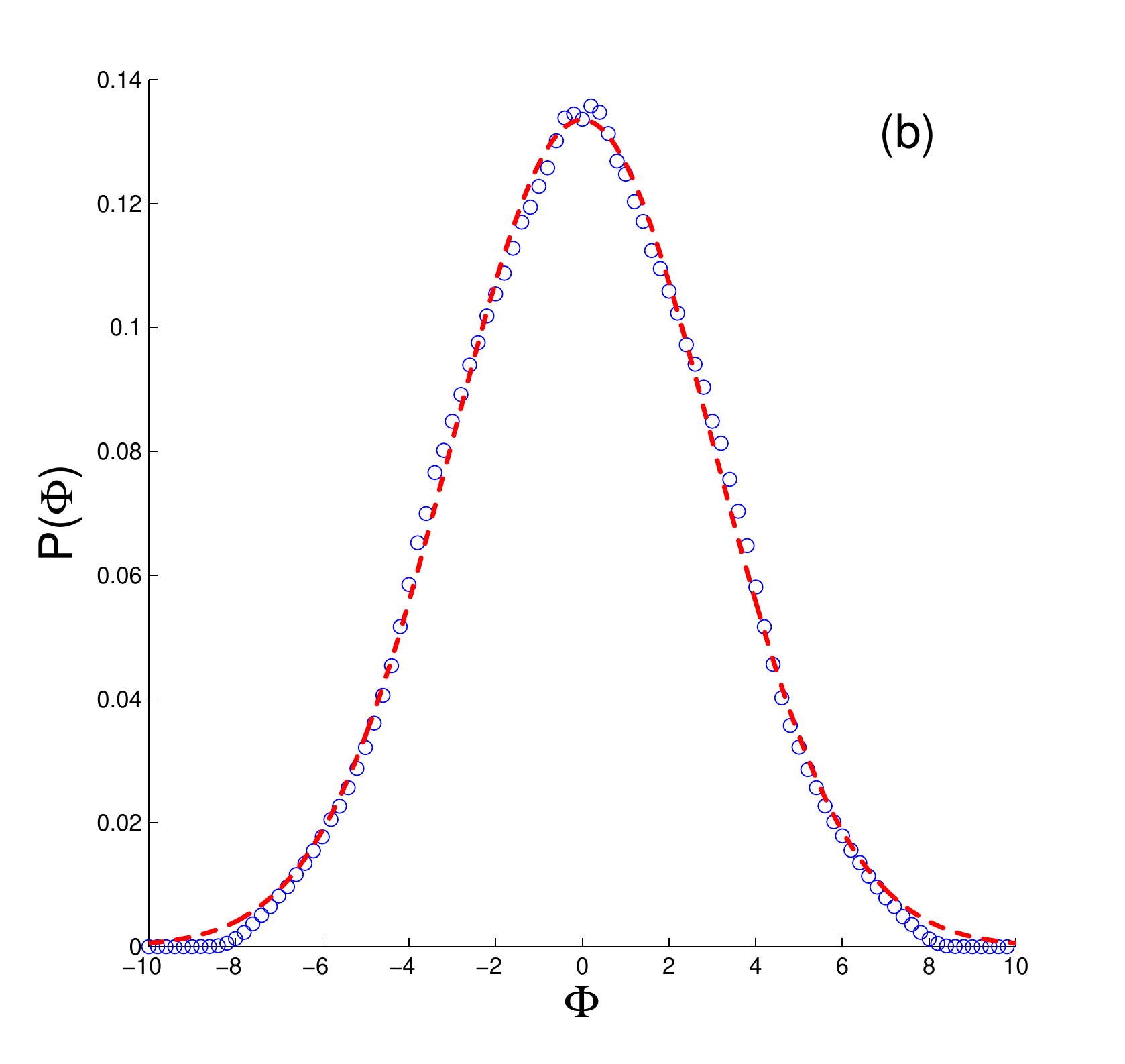}

\includegraphics[scale=0.5]{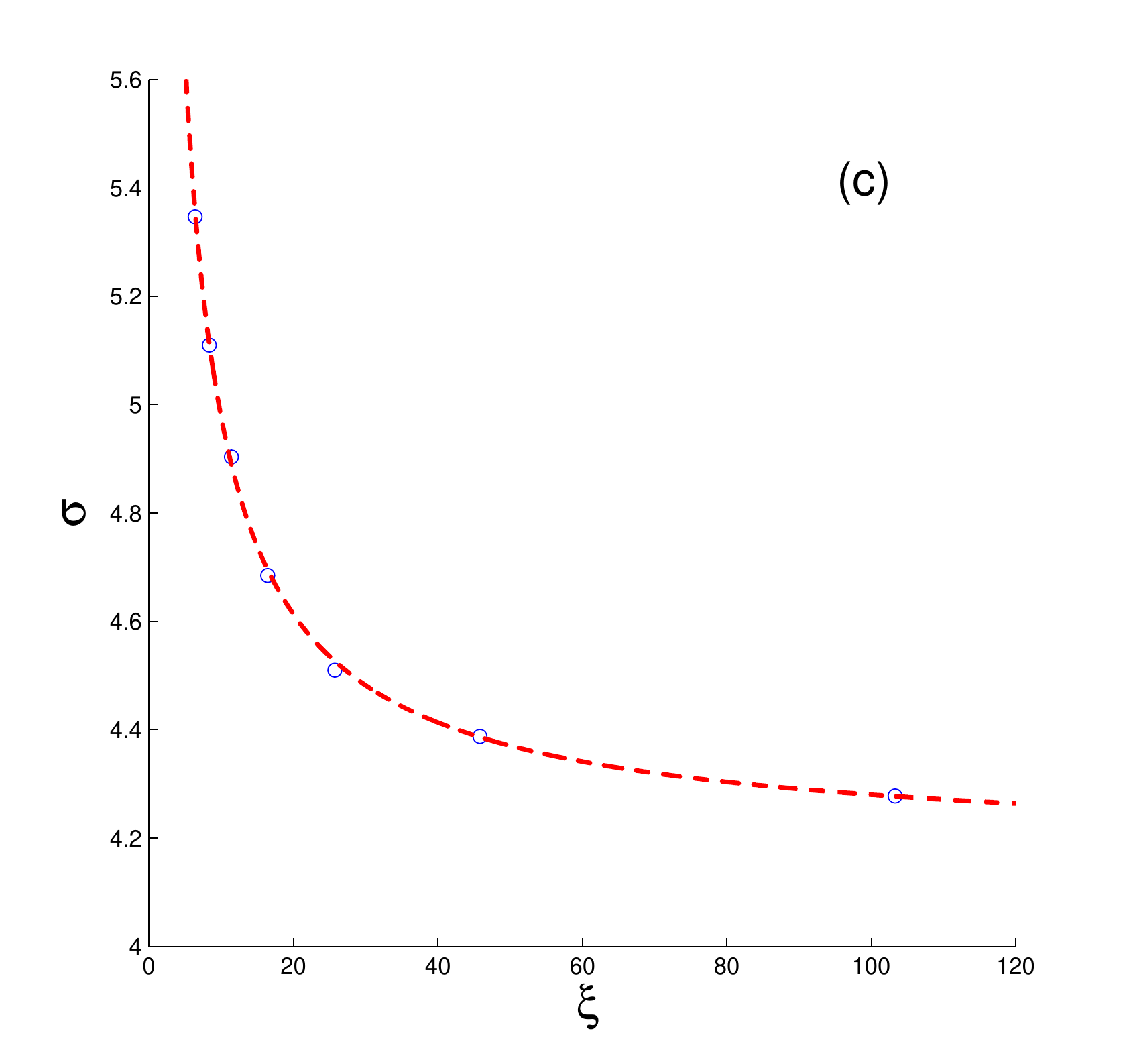}\includegraphics[scale=0.5]{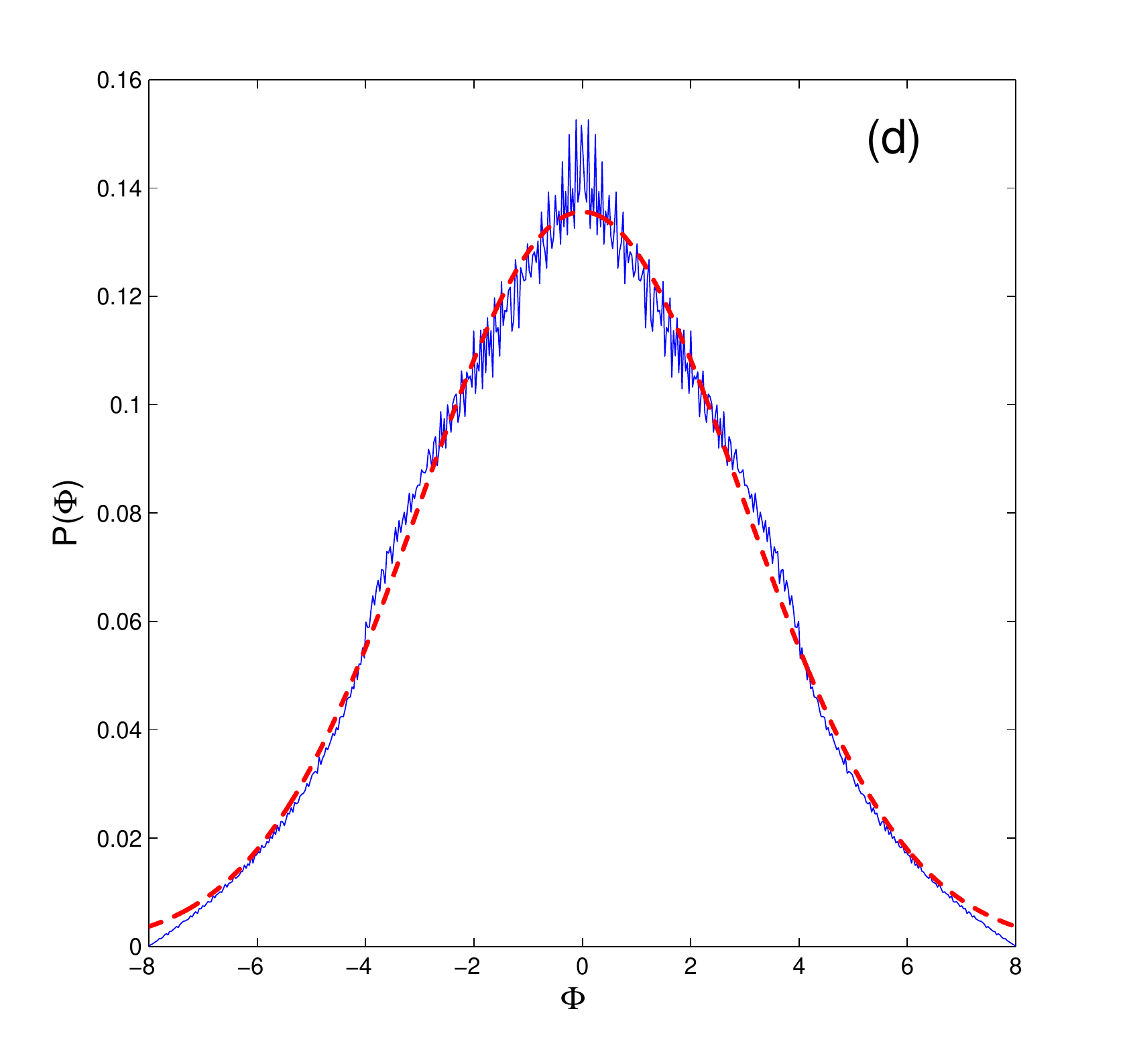}

\caption{\label{fig:All_dis_phi}The PDF of $\Phi$ for various values of weak
disorder strength. (a) blue circles $W=1$ $\left(\xi\approx103\right)$,
green squares $W=1\frac{1}{2}$ $\left(\xi\approx46\right)$ , red
crosses $W=2$ $\left(\xi\approx25\right)$, turquoise dots $W=2\frac{1}{2}$
$\left(\xi\approx16.5\right)$, purple pluses $W=3$ $\left(\xi\approx11.4\right)$,
green stars $W=3\frac{1}{2}$ $\left(\xi\approx8.4\right)$, black
rhombus $W=4$ $\left(\xi\approx6.5\right)$. Number of bins = 1000.
(b) the PDF as a function of $\Phi$ with $W=1$, the red dashed line
is the fit (\ref{eq:fit for gaussian}). (c) $\sigma$ as a function
of $\xi$, the red dashed line it the fit \ref{eq:sigma vs xi}. (d)
The PDF as a function of $\Phi$ for $W=0$, the red dashed line it
the fit (\ref{eq:fit for gaussian}) with values $A=0.1335$ and $\sigma=4.27$.}
\end{figure}

\begin{singlespace}
$\Phi_{m_{1}}^{m_{2},m_{3},m_{4}}=\Phi$ is a combination of 4 eigenenrgies.
In order to calculate the distribution of $\Phi$ one needs to compute
$N^{4}$ of these combinations. To avoid lengthy computations we are
presenting a much smaller number of realizations and use a smaller
lattice size. In this subsection we will present these distribution
for $N_{R}=10$ realizations on a lattice with size $N=128$ for 7
disorder strengths in the range $1\leq W\leq4$, which correspond
to $6.5<\xi<103$. A Gaussian like distribution is found for all values
of $W$, as shown in Fig. \ref{fig:All_dis_phi}a. The form of the
distribution is 
\begin{equation}
P\left(\Phi\right)=Ae^{\frac{-\Phi^{2}}{\sigma^{2}}}\label{eq:fit for gaussian}
\end{equation}
where $A$ is the normalization constant and $\sigma$ is the width
of the gaussian. A fit is presented in Fig. \ref{fig:All_dis_phi}b
for $W=1$ with the values of $A=0.1335\ldots$ and $\sigma=4.278\ldots$
in agreement with \cite{Fishman2009a}. Next we calculated the width
of each Gaussian $\sigma$ as a function of $\xi$ and found 
\begin{equation}
\sigma\left(\xi\right)=b_{1}\cdot\xi^{-b_{2}}+b_{3}\label{eq:sigma vs xi}
\end{equation}
with $b_{1}=5.924\ldots$, $b_{2}=0.865\ldots$ and $b_{3}=4.17\ldots$,
this function is presented in Fig. \ref{fig:All_dis_phi}c. For the
case of very weak disorder, $\xi\gg1$, we see the value of $\sigma$
approaches $\sigma\rightarrow4.17\ldots$, which is in agrement with
the value found for the distribution plotted in Fig. \ref{fig:All_dis_phi}d
where $W=0$.

\end{singlespace}

\begin{singlespace}

\section{Strong disorder}
\end{singlespace}

\begin{figure}[H]
\includegraphics[scale=0.4]{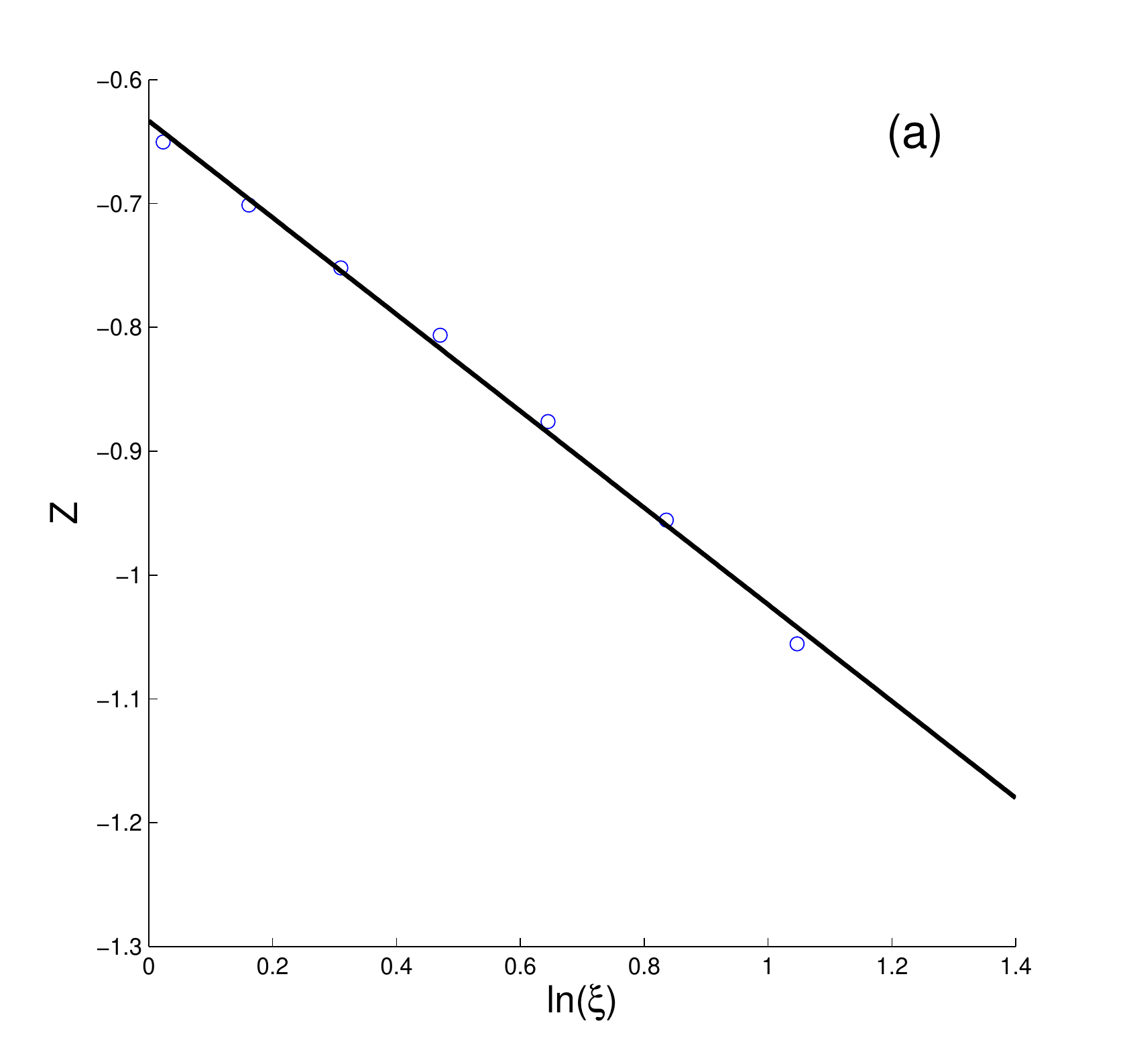}\includegraphics[scale=0.4]{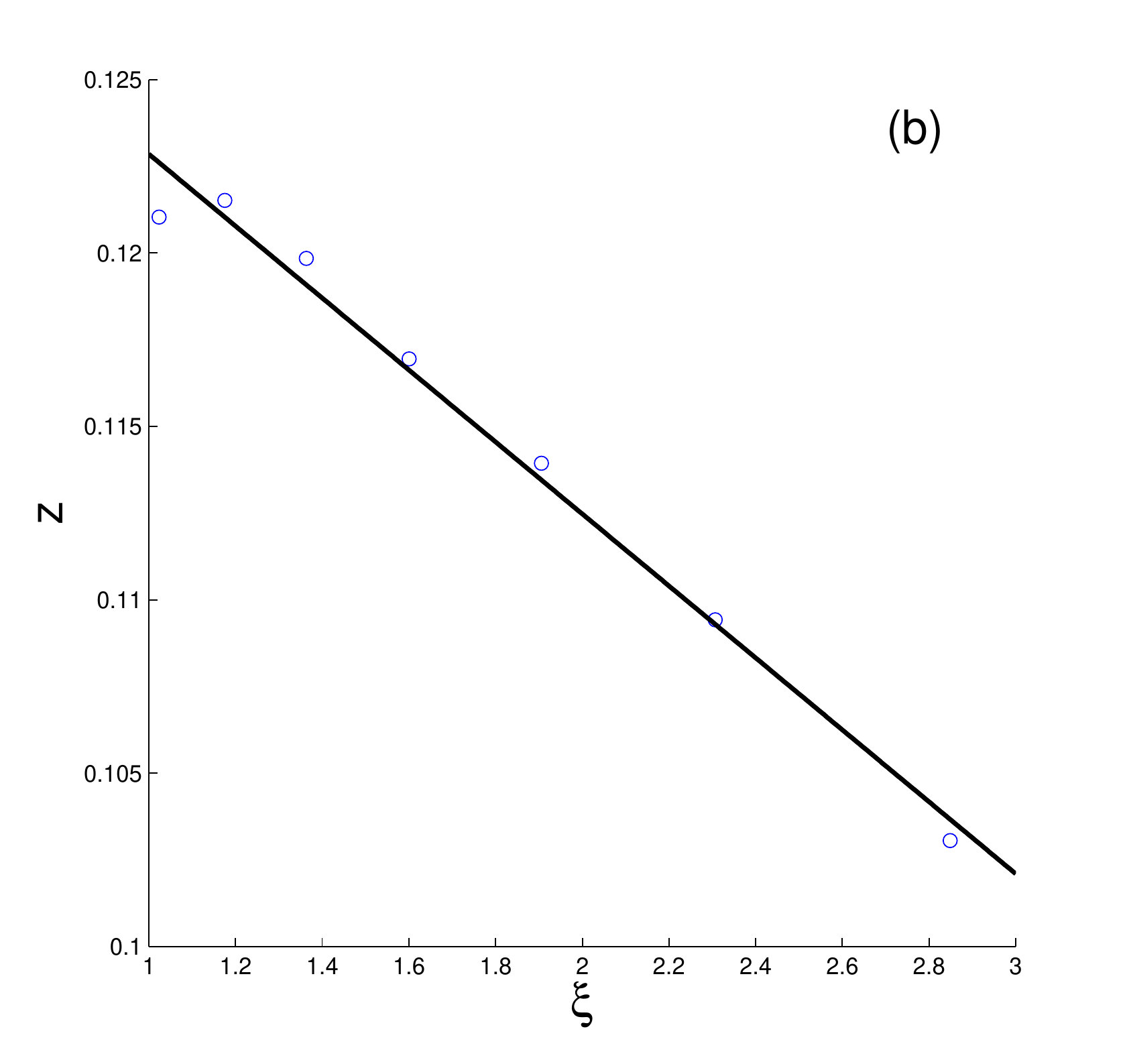}\includegraphics[scale=0.4]{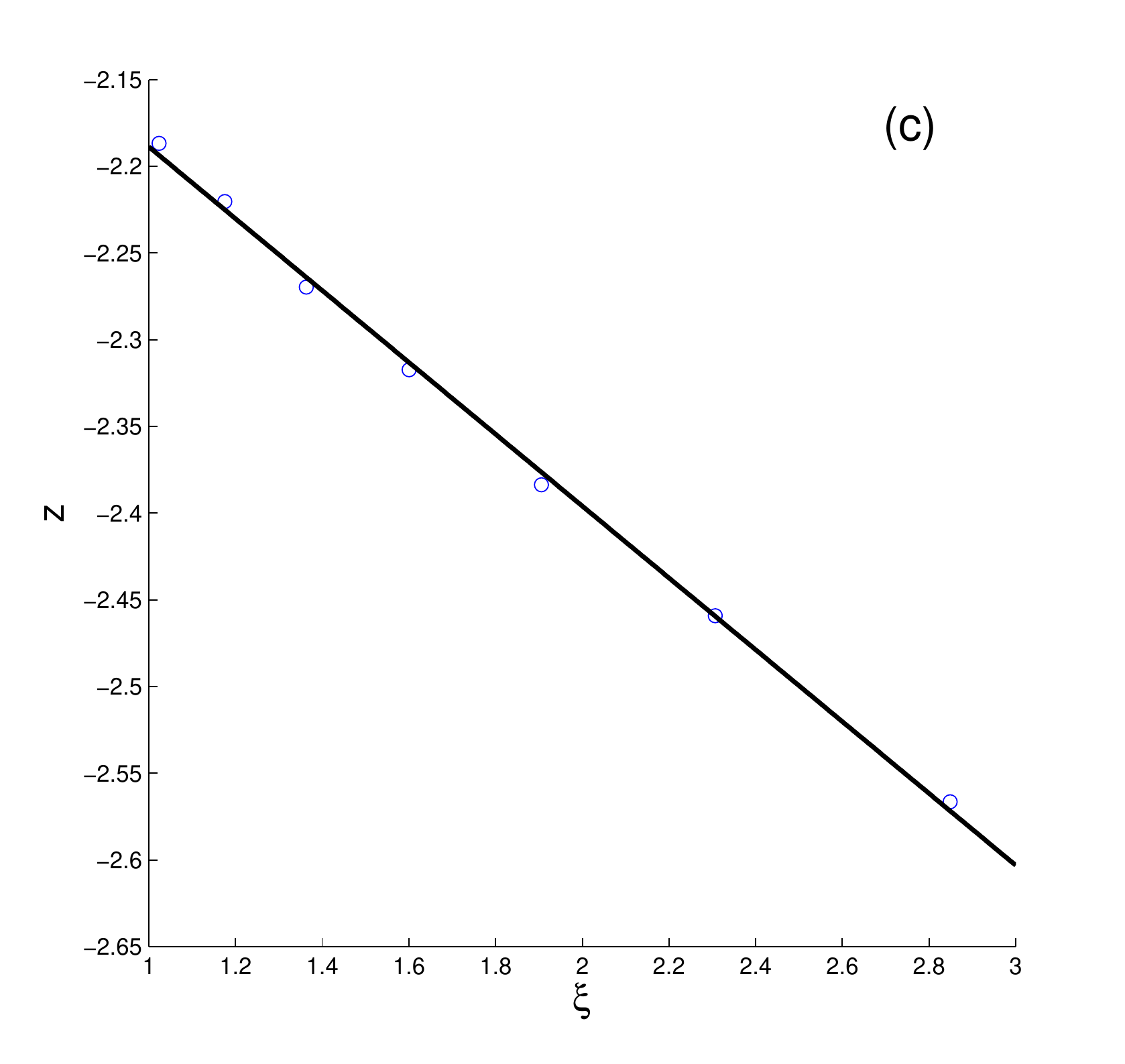}

\caption{\label{fig:(a)-strong disorder mean  V}(a) $z=\ln\left(\left\langle V_{0}\right\rangle \right)$
as a function of $\ln\left(\xi\right)$ with a linear fit $z=-0.39\cdot\ln\left(\xi\right)-0.63$.
(b) $z=\left\langle V_{1}\right\rangle $ as a function of $\xi$
with the fit $z=-0.01\cdot\xi+0.13$. (c) $z=\ln\left(\left\langle V_{2}\right\rangle \right)$
as a function of $\xi$ with a linear fit $z=-0.21\cdot\xi-2$}
\end{figure}

\begin{figure}[H]
\includegraphics[scale=0.5]{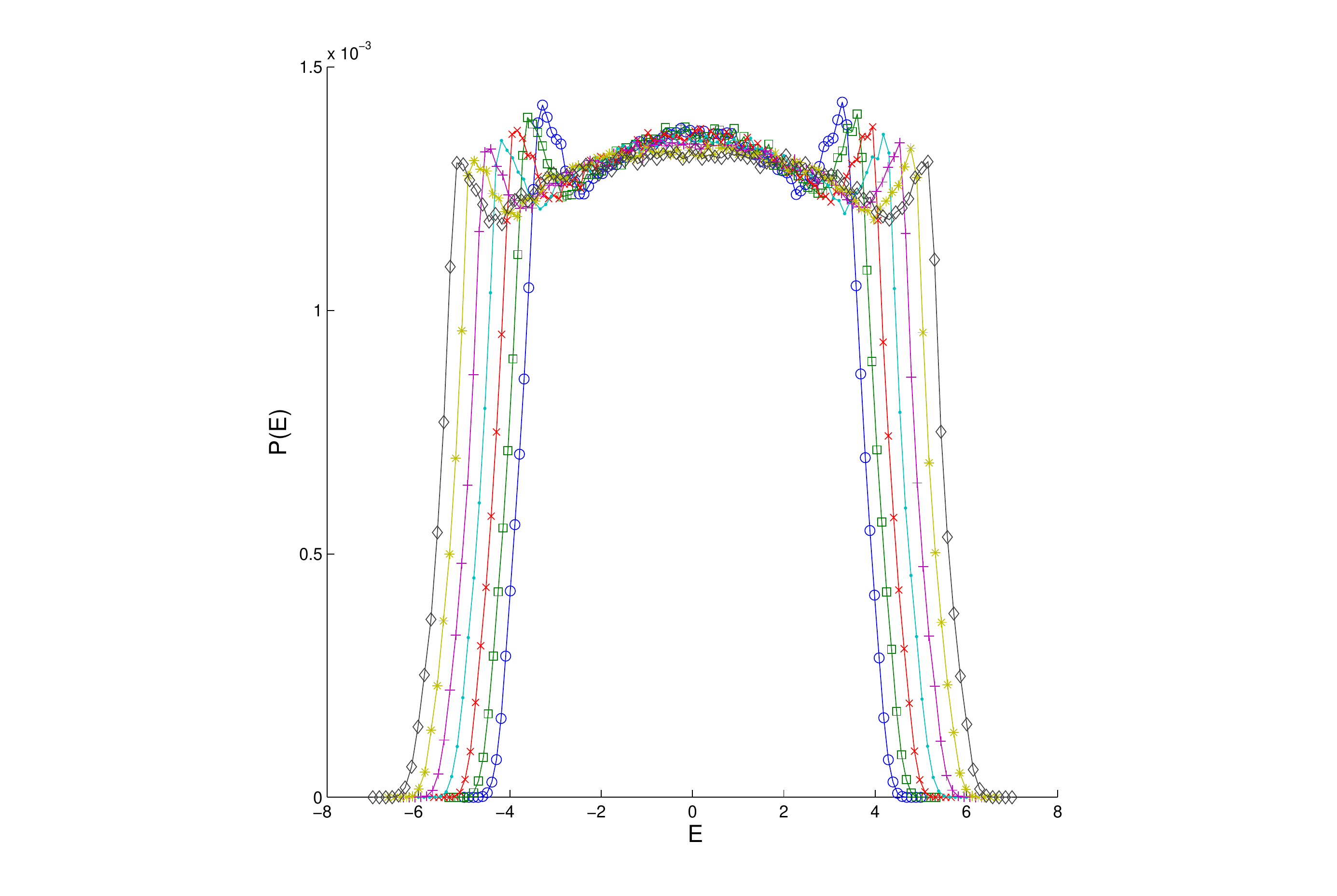}\caption{\label{fig:The-PDF-of_strong disorder _E}The PDF of $E_{n}$ for
various strong disorder strength. blue circles $W=6$ $\left(\xi\approx2.85\right)$,
green squares $W=6\frac{2}{3}$ $\left(\xi\approx2.3\right)$ , red
crosses $W=7\frac{1}{3}$ $\left(\xi\approx1.9\right)$, turquoise
dots $W=8$ $\left(\xi\approx1.6\right)$, purple pluses $W=8\frac{2}{3}$
$\left(\xi\approx1.36\right)$, green stars $W=9\frac{1}{3}$ $\left(\xi\approx1.17\right)$,
black rhombus $W=10$ $\left(\xi\approx1.02\right)$. }
\end{figure}

\begin{figure}[H]
\includegraphics[scale=0.5]{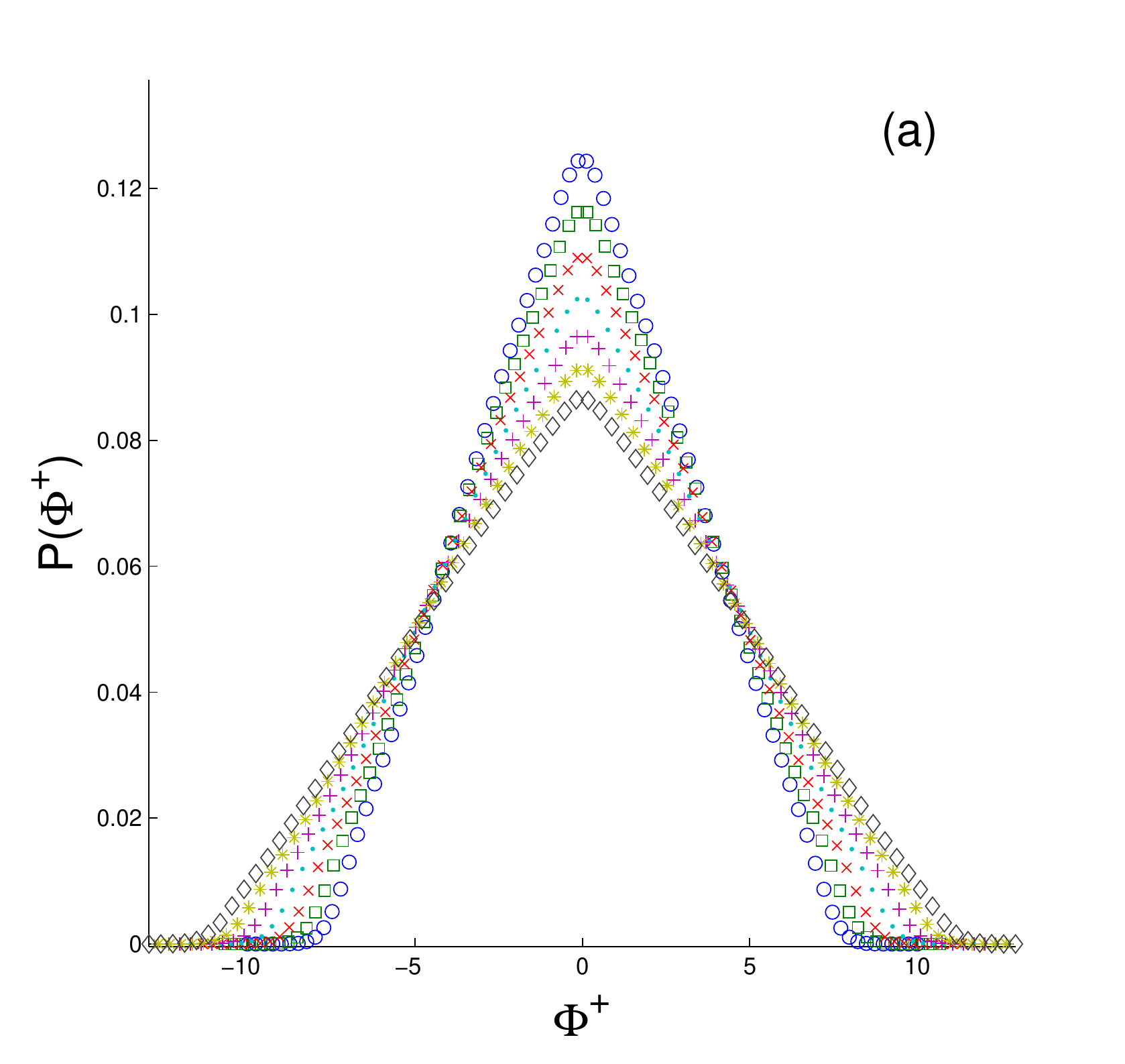}\includegraphics[scale=0.5]{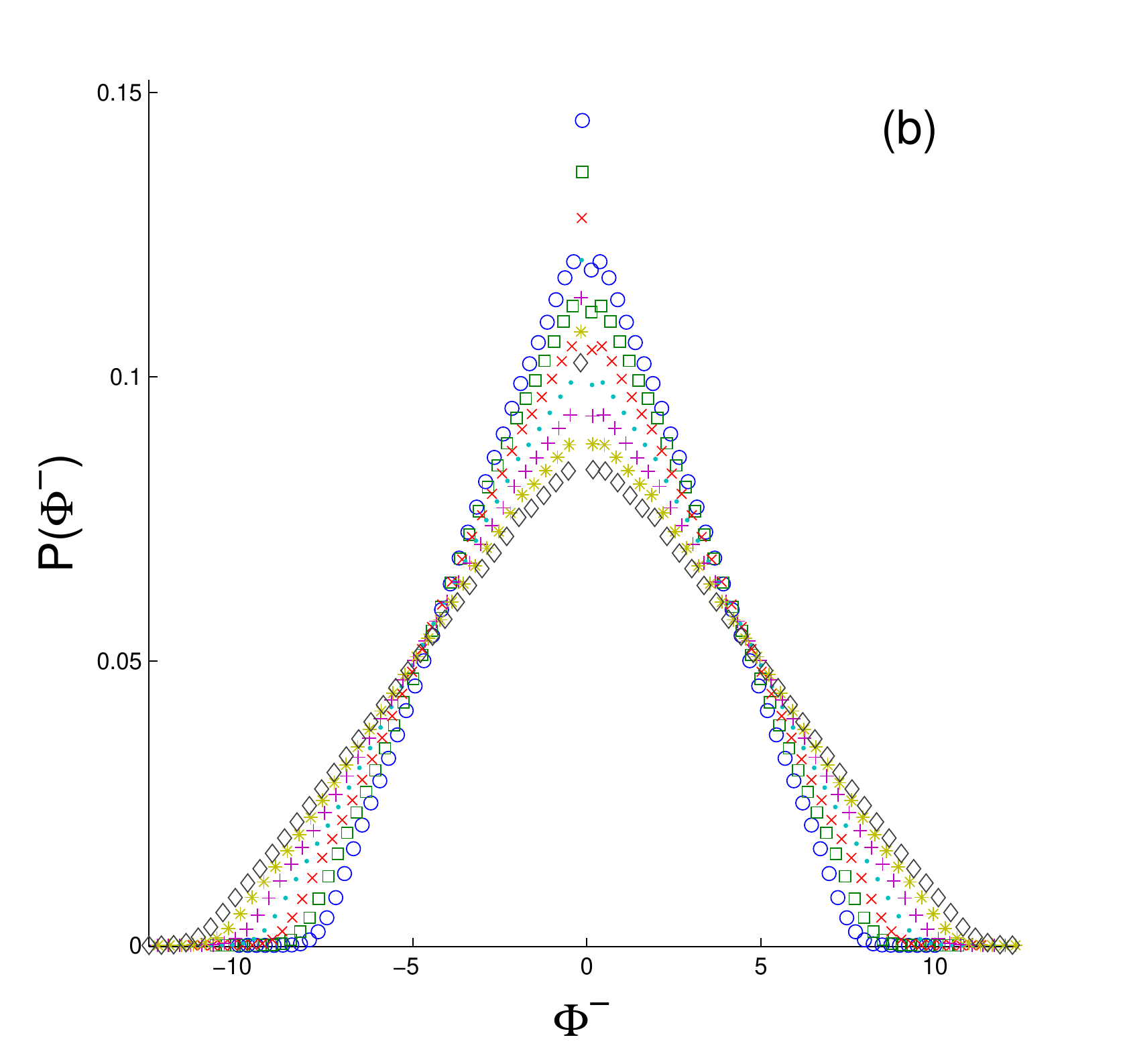}\caption{\label{fig:(a)-The-PDF two energies}Distribution function in the
regime of strong disorder. Symbols are as in Fig. \ref{fig:The-PDF-of_strong disorder _E}.
(a) The PDF of $\Phi^{+}$. (b) The PDF of $\Phi^{-}$.}

\end{figure}

\begin{singlespace}
In the case of strong disorder one does not expect scaling to work
\cite{Shapiro.1988}. Indeed we could not find a scaling distribution
for $V_{0}$ and $V_{1}$ defined in Sec. 2 for the regime of strong
disorder. Their averages scale with different powers of $\xi$ as
is clear from Fig. \ref{fig:(a)-strong disorder mean  V}. The distribution
of the $E_{n}$ exhibits a maximum near $E=0$ as one can see from
Fig. \ref{fig:The-PDF-of_strong disorder _E}, while for weak disorder
a minimum is found there (compare Fig. \ref{fig:(a)-The-distribution_weak_enegies}
to Fig. \ref{fig:The-PDF-of_strong disorder _E}). The distributions
of $\Phi^{+}$and $\Phi^{-}$ presented in Fig \ref{fig:(a)-The-PDF two energies}
exhibit a linear dependence on the values of $\Phi^{+}$ and $\Phi^{-}$
respectively. The distribution of $\Phi$ is similar to the one found
for weak disorder, Fig. \ref{fig:All_dis_phi}b fits even better gaussian
distribution.

\end{singlespace}

\begin{singlespace}

\section{Summary and Discussion}
\end{singlespace}

\begin{singlespace}
In this work the distribution function of some overlap sums $V_{m_{1}}^{m_{2},m_{3},m_{4}}$
of (\ref{Cs}) were calculated. In some cases of weak disorder it
was demonstrated that scaling holds. In particular it was shown that
for weak disorder (or large localization length $\xi)$ the distribution
functions of $V_{0}^{0,0,0}$, $V_{0}^{0,1,1}$ , $V_{0}^{0,0,1}$
are functions of these variables and of $\xi$ via one scaling variable,
$y_{0}=\xi V_{0}^{0,0,0}$; $y_{1}=\xi V_{0}^{0,1,1}$ and $y_{2}=\xi^{3/2}V_{0}^{0,0,1}$
see (\ref{eq:scaling IPR}), (\ref{eq:scaling V_1}) and (\ref{eq:scaling variable y_2}).
The distribution function of $y_{0}$ was calculated numerically (\ref{eq:universal_IPR}).
We could not find a scaling function for $V_{0}^{0,1,2}$ and $V_{0}^{1,2,3}$.
In addition to the fundamental interest, the scaling function can
be extremely useful for the case when $\xi$ is very large. It enables
to obtain the distribution of $V_{m_{1}}^{m_{2},m_{3},m_{4}}$ for
regimes where numerical calculations require a basis of size that
is beyond the available computer resources. Also the averages and
variances of the $V_{m_{1}}^{m_{2},m_{3},m_{4}}$ were computed for
some $\left\{ m_{i}\right\} .$ In most cases of weak disorder we
analyzed, these scale simply with $\xi$ as one could guess from the
distribution functions of the scaling variable. In some cases the
averages are exponential in $\xi$ (see Fig. \ref{fig:(a) mean-xi_3 indices}).

The distribution functions of combinations of energies of the Anderson
model were studied as well. A difference between $\Phi^{+}=E_{n}+E_{m}$
and $\Phi^{-}=E_{n}-E_{m}$ was found near $\Phi^{+}=0$, $\Phi^{-}=0$.
It is a signature of level repulsion. The distribution of $\Phi_{m_{1}}^{m_{2},m_{3},m_{4}}=E_{m_{1}}+E_{m_{2}}-E_{m_{3}}-E_{m_{4}}$
was found to be gaussian and the dependence of the variance on the
localization length was computed (\ref{eq:sigma vs xi}).

Some results for the various distributions were calculated in the
regime of strong disorder. Results that are typically different from
ones found for weak disorder were obtained. 

The research can be continued in two directions. The distribution
can be calculated numerically for additional cases and divided into
classes with different characteristic properties. Also better analytical
insights are called for.

This work was motivated by the effective noise theory for Nonlinear
Schrödinger Equation (NLSE) in a random potential \cite{Michaely2012,Flach2009,Skokos2009}
and the results of this work are of great importance for this theory.
However the results may be of interest by themselves.
\end{singlespace}

\begin{singlespace}

\paragraph{Acknowledgment}
\end{singlespace}

\begin{singlespace}
We would like to thank Y. Krivolapov for detailed discussions, extremely
valuable technical detailed help and for extremely critical reading
of the menuscript . This work was partly supported by the US-Israel
Binational Science Foundation (BSF), by the Minerva Center of Nonlinear
Physics of Complex Systems, and by the Shlomo Kaplansky academic chair. 

\bibliographystyle{unsrt}
\bibliography{C:/Users/Erez/Documents/Papers/NLSE}
\end{singlespace}

\end{document}